\documentclass{article}

\usepackage{arxiv}

\renewcommand{\shorttitle}{Detecting Internal Contradictions in Privacy Policies}

\usepackage[utf8]{inputenc}
\usepackage[T1]{fontenc}
\usepackage{microtype}

\usepackage{array}
\usepackage{booktabs}
\usepackage{graphicx}

\usepackage{amsmath}
\usepackage{amssymb}

\usepackage{tikz}
\usetikzlibrary{shapes.geometric, arrows.meta, positioning, calc, fit, backgrounds}

\usepackage{natbib}
\usepackage{hyperref}

\begin{document}

\title{Privacy Washing: Detecting Internal Contradictions in Privacy Policies}

\author{
Thomas Brackin \\
Varitas (varitas.io) \\
Seattle, WA, USA \\
\texttt{thomas@varitas.io}
}

\date{August 31, 2026}

\maketitle

%%
%% Abstract
\begin{abstract}
Privacy policies may contain internal contradictions in which commitments are undermined by practices documented elsewhere in the same policy. This paper operationalizes this phenomenon, \emph{privacy washing}, through a four-stage automated pipeline: statement extraction, compatibility filtering and natural language inference (NLI) screening, multi-model judge verification, and thematic analysis. Contradictions are confirmed by majority vote of a three-model large language model (LLM) panel. Applied to two corpora of website privacy policies, 123 collected in 2026 (the Open Privacy Policy Taxonomy corpus, OPPT) and 115 collected in 2015 (OPP-115), the pipeline finds the same category patterns recurring across the 11-year gap, with contradictions involving third-party sharing accounting for the majority of confirmed cases in each corpus in the primary runs, consistent with structural factors in policy composition rather than necessarily intentional deception. At least one panel-confirmed contradiction appears in 12.2\% of OPPT companies (15/123; 9.8\%, 12/123, in the reproducible subset that excludes legacy pairs) and 36.5\% of OPP-115 companies (42/115). A stability re-run seven months later, with a fully separated configuration (new extraction models, judge models from three Chinese providers absent from both corpora, matched filters, and no similarity threshold at judge submission), reproduces the OPPT prevalence under the original protocol (13.0\% vs.\ 12.2\%), shows that pairs below the similarity threshold confirm at rates of the same order as those above it (raising prevalence to 20.3\% and 40.9\%), and shows that the third-party majority is panel-sensitive while the recurrence of the same category pairs is not. Two caveats govern all reported figures. First, panel verdicts have not been validated against human expert judgment, so precision is unknown, and the pipeline's aggressive filtering makes the prevalence figures lower bounds. Second, the two primary corpora runs used different filter configurations, so the difference in prevalence between them is not interpretable as a corpus or era effect; the matched-configuration re-run reduces the gap to roughly twofold but does not eliminate it.
\end{abstract}


\section{Introduction}

A company's privacy policy may promise ``we do not sell your personal information'' in one section while describing extensive third-party data sharing in another. Both statements appear in the same document, governing the same users. A reader who encounters the reassuring commitment may not reach the contradictory practice described several sections later. Prior work establishes that most users skip policies or navigate by section headings rather than reading comprehensively~\citep{obar2020biggest,vu2007users}, which is the condition under which a commitment and a distant contradicting practice would not be read together. We analyze this pattern under the term \emph{privacy washing}, which has prior use in industry commentary~\citep{linton2021privacywashing} and in academic critique of privacy marketing~\citep{cirucci2024privacywashing}; we adapt it to within-policy contradictions and provide, to our knowledge, its first computational operationalization. Whether readers actually form expectations based on such commitments is an empirical question beyond the scope of this paper; our contribution is to detect and characterize the structural pattern.

The notice-and-choice framework underpins privacy regulation worldwide, from the EU's General Data Protection Regulation (GDPR)~\citep{gdpr2016} to the California Consumer Privacy Act (CCPA)~\citep{ccpa2018}. This framework assumes that policies provide meaningful notice enabling informed decisions. Extensive research has shown this assumption frequently fails: policies are too long~\citep{mcdonald2008cost}, too complex~\citep{fabian2017readability}, and structurally fragmented~\citep{brackin2026jurisdiction}, and the consent paradigm itself faces structural limitations~\citep{solove2013introduction,cranor2012necessary}. These findings have motivated simplified disclosures~\citep{kelley2009nutrition}, layered notices~\citep{schaub2015design}, and machine-readable formats.

Yet these reforms address only one side of the problem. Readability improvements assume that a policy's content is internally consistent. A user who successfully reads and understands a commitment should be able to rely on it. Our findings identify candidate cases where this assumption fails. Across two corpora spanning 11 years (123 companies in OPPT, collected in 2026, and 115 in OPP-115, collected in 2015), we identify patterns where commitments are contradicted by practices documented elsewhere in the same policy. A policy promising ``we do not sell your personal information'' while describing disclosure of identifiers and commercial information to advertising partners creates conditions under which a reader who encounters the commitment may form expectations that the practices contradict. This pattern is consistent with the 2022 Sephora settlement, in which the California Attorney General determined that sharing data with advertising partners constituted ``sale'' under the CCPA~\citep{sephora2022settlement}.

The analogy with greenwashing~\citep{delmas2011greenwashing} is deliberate: just as a company may advertise a minor environmental initiative while maintaining harmful practices, a privacy policy may feature reassuring commitments undermined by its own fine print. The contradiction is pragmatic rather than purely logical: both sides may be individually defensible, but together they create a misleading impression.

Detecting privacy washing computationally presents distinct challenges. Prior contradiction detection systems for privacy policies, notably PolicyLint~\citep{andow2019policylint} and PoliGraph~\citep{cui2023poligraph}, identify logical negation within structured data-collection tuples (``we collect X'' vs.\ ``we do not collect X''). These systems are not designed to capture pragmatic tension between a broad reassurance and a specific practice, a pattern that symbolic tuple inversion cannot express. Detecting this class requires distinguishing between commitments and practices, pairing them directionally, and applying verification beyond surface-level textual similarity.

This paper makes three contributions:

\begin{enumerate}
    \item We define \emph{privacy washing} and give it an explicit operationalization (the judge prompt of Appendix~\ref{appendix:judge}), identifying where the operationalization and the conceptual definition diverge. We identify recurring contradiction patterns through empirical analysis of 238 companies across two corpora, the Open Privacy Policy Taxonomy (OPPT) corpus and OPP-115, spanning diverse industries and regulatory eras.

    \item We present a four-stage automated pipeline: consensus extraction of typed atomic statements with structured metadata (subject, aspect, scope, qualifiers) by three large language models (LLMs), category and metadata compatibility filtering followed by a semantic similarity pre-filter, natural language inference (NLI) screening, and multi-model judge verification as the final precision layer.

    \item We provide a cross-corpus analysis of two corpora collected 11 years apart, OPPT (123 companies, 2026) and OPP-115 (115 companies, 2015), showing that the same small set of first-party and third-party category pairs recurs in both under a fixed operationalization, with third-party sharing contradictions the most frequent in each primary run, a majority that a re-run with a different judge panel shows to be panel-sensitive (Section~\ref{sec:stability}), suggesting structural factors in policy composition. Because the two primary runs used different filter configurations, we do not interpret the prevalence difference between the corpora.
\end{enumerate}

Amid the caveats that accompany these results, what the paper establishes affirmatively is a reproducible operational definition of privacy washing, a released artifact set (pipeline, intermediate outputs, and judge verdicts) that enables community validation, and a set of category patterns that recur across an 11-year gap under that fixed operationalization. The OPPT corpus was collected for a companion study on jurisdiction-siloed disclosure~\citep{brackin2026jurisdiction}, which examined how policy structure can undermine notice. This paper examines a complementary failure: how policy content can undermine notice through internal contradiction.

\section{Background and Related Work}

\subsection{Privacy Policy Analysis}

Automated analysis of privacy policies has been an active research area since early machine learning approaches~\citep{costante2012machine}. The creation of the OPP-115 corpus~\citep{wilson2016creation} established a standardized annotation scheme for 115 website privacy policies across 10 categories. Unsupervised alignment methods revealed structural regularities in policy text~\citep{ramanath2014unsupervised}. OPP-115 has served as training data for classifiers including Polisis~\citep{harkous2018polisis}, and has been complemented by later resources such as the PrivacyQA question-answering corpus~\citep{ravichander2019question} and the PrivaSeer search engine and corpus~\citep{srinath2021privaseer}, which extended policy analysis to web scale through domain-adapted language models.

A parallel line of work has examined policy-to-code consistency. MAPS~\citep{zimmeck2019maps} compared Android app privacy policies against actual app behavior at scale, finding widespread inconsistencies. PoliCheck~\citep{andow2020policheck} extended this with entity-sensitive flow-to-policy analysis, detecting that up to 42.4\% of apps exhibited behavior inconsistent with their stated policies. ATLAS~\citep{jain2023atlas} applied 32 classifiers to compare Apple privacy labels against actual policy text, and found at least one possible discrepancy between label and policy in 88\% of apps that provided both an accessible policy and a label.

Large-scale longitudinal resources also exist: \citet{amos2021privacy} curated over one million privacy policy snapshots from 130,000 websites spanning two decades, and \citet{srinath2021corpus} released a corpus of over one million contemporary policies. We compare two smaller purpose-annotated corpora in this paper because our method currently uses per-segment annotations to guide statement typing (Section~\ref{sec:methodology}), which the larger corpora lack; applying it to these larger corpora is a natural next step (Section~\ref{sec:future}). \citet{ravichander2021breaking} surveyed how NLP can benefit consumer privacy while cautioning against premature deployment of policy-analysis systems. More recently, \citet{rodriguez2024llm} demonstrated that large language models can annotate privacy policies at scale with accuracy comparable to or exceeding prior approaches on established benchmarks. A comprehensive ACM Computing Surveys review of privacy policy NLP~\citep{javed2024review} found that mismatches between stated policies and actual data practices are the least-studied area in the field, addressed by only 10.9\% of papers, while classification dominates.

\subsection{Contradiction Detection in Privacy Policies}

The most closely related line of work is contradiction detection within privacy policies. PolicyLint~\citep{andow2019policylint} pioneered this direction using symbolic NLP to extract data-collection four-tuples (actor, action, data\_object, entity) and detect nine contradiction types arising from subsumptive relationships between terms, five of which are logical contradictions and four narrowing definitions. Applied to 11,430 Android app policies, PolicyLint found contradictions in 14.2\% of policies. PoliGraph~\citep{cui2023poligraph} advanced this approach using knowledge graphs, extracting 40\% more policy statements than PolicyLint and enabling document-level consistency analysis. \citet{okoyomon2019ridiculousness} documented contradictions between app privacy policies and observed data flows, and among policy statements themselves, arguing that such contradictions undermine notice and consent; their analysis relies on flow-to-policy comparison rather than within-document commitment-practice pairing.

PurPliance~\citep{bui2021purpliance} introduced purpose-awareness to contradiction detection, identifying inconsistencies in how data-usage purposes are described across policy sections, and found contradictions in 18.14\% of analyzed policies. PolicyLR~\citep{hooda2024policylr} brought formal logic representation to policy analysis, compiling policies into machine-readable logic formulas that support automated reasoning about consistency and compliance. A precursor line in requirements engineering detects conflicts over formally specified data flows: \citet{breaux2014eddy} introduced the Eddy language for analyzing conflicting privacy requirements extracted from policies, and \citet{breaux2015detecting} detected repurposing and over-collection across multi-party requirements specifications. These systems reason over specified data-flow requirements, whereas our target is pragmatic contradiction between natural-language statements.

These systems share a common scope restriction: they detect logical contradictions (direct negation within structured tuples) but do not target pragmatic contradictions where a broad reassurance is undermined by a specific practice. ``We do not sell your personal information'' paired with ``we disclose identifiers and commercial information to advertising partners'' is not a logical negation of any data-collection tuple. The contradiction lies in the gap between the impression the commitment creates and the practice the policy documents elsewhere, a pragmatic tension that symbolic methods are not designed to detect.

Outside the privacy domain, the foundational NLP typology of contradiction is due to \citet{demarneffe2008finding}, who distinguished seven types: antonymy, negation, and numeric mismatch (detectable from surface features) versus factive/modal, structural, lexical, and world-knowledge contradictions (requiring deeper comprehension). These types are semantic in character: the incompatibility is recoverable from the truth conditions of the two statements. The world-knowledge category comes closest to the commitment-practice tension we target, but it still presumes the two statements cannot both be true. Our target sits at the boundary of this category: both statements can be individually true and propositionally consistent while the practice undermines the impression the commitment creates; our ``pragmatic contradiction'' names this gap. ContraDoc~\citep{li2024contradoc} addressed self-contradiction in documents using LLMs, demonstrating that contradiction detection benefits from document-level context. ContraDoc does not distinguish between statement types and is not tailored to the commitment-versus-practice structure characteristic of privacy policies. CLAUDETTE~\citep{lippi2019claudette} detected potentially unfair clauses in terms of service using machine learning but targets unfairness rather than internal contradiction.

\subsection{Natural Language Inference for Legal Text}

Natural language inference has been applied to legal and regulatory text with mixed success. ContractNLI~\citep{koreeda2021contractnli} established a dataset of 607 annotated contracts for document-level NLI, demonstrating that NLI models can reason about contractual entailment when provided with domain-appropriate training data. EXCLAIM~\citep{ikhwantri2025exclaim} formulated compliance detection as multi-hop NLI over the claim-argument-evidence structure of assurance cases, demonstrated on GDPR requirements.

A persistent challenge is domain transfer. NLI models trained on SNLI~\citep{bowman2015snli} and MultiNLI~\citep{williams2018multinli} achieve above 90\% accuracy on general benchmarks but degrade markedly on legal text. A DeBERTa-large model fine-tuned on SNLI and MultiNLI achieved only 66.34\% accuracy on the statute law entailment task of the COLIEE 2023 shared task~\citep{kim2024legal,goebel2024coliee}. NLI models are also known to exploit annotation artifacts and lexical overlap heuristics rather than learning genuine inference~\citep{gururangan2018annotation,mccoy2019right}, raising concerns about their reliability on specialized domains where such shortcuts may not hold.

No existing NLI dataset targets within-policy contradictions. ContractNLI addresses contract-level inference and EXCLAIM addresses policy-regulation compliance, but neither examines whether a single document contradicts itself. This gap motivates using NLI as a screening signal rather than a final arbiter, with multi-model verification providing precision that NLI alone cannot achieve.

\subsection{The Greenwashing Analogy}

The term \emph{privacy washing} draws on the established concept of greenwashing: corporate claims that misrepresent environmental commitments or practices. \citet{delmas2011greenwashing} identified the structural drivers of greenwashing in corporate sustainability, and the TerraChoice ``Seven Sins of Greenwashing'' framework~\citep{terrachoice2010sins} provided an influential taxonomy. Automated greenwashing detection in text has become an active NLP research area, spanning claim detection, specificity assessment, and deception identification~\citep{calamai2025greenwashing}. The nearest methodological analogue to our task is cheap-talk detection in corporate climate disclosures, where \citet{bingler2022cheap} used a domain-adapted language model to measure, at scale, the gap between reassuring corporate language and substantive commitment in a regulated disclosure genre. The greenwashing literature has evolved beyond requiring proof of intent. \citet{bowen2014greenwashing} demonstrated that greenwashing often arises from organizational complexity rather than calculated deception, and TerraChoice characterizes greenwashing as ``exaggeration rather than falsehood,'' a framing that does not presume deliberate intent. Our use of ``privacy washing'' follows this structural tradition, identifying measurable patterns in policy text without attributing motive to authors.

In the privacy domain, several scholars have identified analogous phenomena. \citet{waldman2020false} argued that privacy law creates ``false promise'' through compliance theater, where companies satisfy legal requirements without providing meaningful privacy protection. \citet{soghoian2011privacy} used ``privacy theater'' to describe corporate privacy practices that appear protective but lack substance. \citet{cirucci2024privacywashing} applied ``privacy-washing'' to marketing that conflates security features with privacy protection, and \citet{lestyan2025anonymity} identified ``anonymity-washing,'' the overstatement of the anonymity of ostensibly anonymized data. Empirical work in usable privacy documents adjacent gaps between stated and delivered privacy choices: \citet{cranor2016largescale} found internal inconsistencies in U.S. financial institutions' standardized privacy notices, and \citet{habib2019empirical,habib2020scavenger} showed that opt-out and deletion choices described in policies are frequently unusable in practice. Our usage extends this family of terms from marketing and anonymization claims to contradictions within the policy document itself. Privacy washing also relates to dark patterns in user interfaces~\citep{mathur2019dark,gray2018dark,luguri2021dark,nouwens2020dark}, though it operates through policy text rather than interface design.

\citet{belcheva2023understanding} found that privacy policies increasingly use ``pacifying phrases'' alongside increasing vagueness, supporting the hypothesis that reassuring language and substantive hedging coexist systematically.

Despite these conceptual foundations, no prior work has operationalized commitment-versus-practice contradiction in privacy policies as a computationally detectable phenomenon. Our contribution formalizes privacy washing through a precise definition, detection pipeline, and empirical measurement across two corpora.

\section{Defining Privacy Washing}
\label{sec:definition}

\subsection{Definition}

We define \emph{privacy washing} as the presence within a single privacy policy of one or more commitments (promises, reassurances, or stated limitations on data practices) that are contradicted by practices (descriptions of actual data handling) documented elsewhere in the same policy. Following the greenwashing literature's structural tradition~\citep{terrachoice2010sins,bowen2014greenwashing}, we identify patterns in policy text without attributing motive to authors.

We acknowledge that ``washing'' carries connotations of deliberate deception that our findings do not establish. The doctrinal basis for the term is the Federal Trade Commission's (FTC) ``net impression'' standard (discussed below), under which a technically true statement can mislead through the overall impression it creates, without any showing of intent; privacy washing names this condition arising inside a single policy document. The greenwashing literature provides a secondary parallel: it has established that ``washing'' can describe structural patterns arising from organizational complexity rather than calculated deception~\citep{bowen2014greenwashing}, and the term captures the directional relationship between reassuring commitments and contradicting practices that ``inconsistency'' does not. We take no position on cause: the pipeline cannot distinguish deliberate drafting choices from accretive maintenance for any individual case, and we present the structural factors in Section~\ref{sec:structural} as plausible mechanisms rather than as findings.

The greenwashing analogy is imperfect in one respect: greenwashing typically involves a gap between public claims and private behavior, whereas privacy washing involves contradictions within a single public document; both the commitment and the contradicting practice are disclosed. This within-document structure makes privacy washing amenable to automated detection, and it is also why the net impression standard, which concerns the impression a disclosure itself creates, fits the phenomenon more closely than the greenwashing parallel does.

We use ``contradiction'' to refer to pragmatic contradiction unless otherwise specified; the distinction from logical contradiction (the focus of prior systems like PolicyLint) is central to our contribution. Our pipeline detects textual conditions consistent with privacy washing: commitment-practice pairs that exhibit pragmatic tension.

Three properties distinguish privacy washing from other forms of policy inconsistency:

\begin{enumerate}
    \item \textbf{Within-document}: Both the commitment and the contradictory practice appear in the same policy, distinguishing privacy washing from code-versus-policy inconsistency~\citep{andow2020policheck,zimmeck2019maps}.

    \item \textbf{Directional}: A commitment is undermined by a practice, not merely any pair of inconsistent statements. Two practice statements in tension (``we share data with advertising partners'' and ``we share only aggregated data'') are not evaluated; the pipeline evaluates only COMMITMENT-to-PRACTICE pairs, in that direction.

    \item \textbf{Pragmatic}: The contradiction is pragmatic rather than logical. ``We only share data when legally required'' paired with ``we share data with advertising partners'' is a pragmatic contradiction (the first statement creates an impression of minimal sharing that the second undermines), not a logical negation detectable by symbolic methods.
\end{enumerate}

We distinguish privacy washing from \emph{commitment avoidance}, where companies avoid contradictions by making few commitments rather than by limiting practices; Sections~\ref{sec:commitment_avoidance} and~\ref{sec:vagueness} examine this distinction in detail.

We ground pragmatic contradiction in two frameworks. Speech act theory~\citep{searle1969speech} treats privacy commitments as \emph{commissive speech acts} creating normative obligations; a documented practice that violates such a commitment produces a \emph{felicity failure}. The Cooperative Principle~\citep{grice1975logic} supports the expectation that policies truthfully represent material practices, though its assumption of cooperative conversation may not fully hold for privacy policies. The impression created by qualified commitments such as ``we only share data when legally required'' is a scalar implicature~\citep{horn1984taxonomy,levinson2000presumptive}: the qualifier implicates that stronger sharing claims do not hold.

Contextual integrity~\citep{nissenbaum2004contextual,nissenbaum2010privacy} offers a complementary account in which privacy expectations are governed by context-relative informational norms; \citet{shvartzshnaider2019going} operationalized this framework for privacy policy analysis by annotating policy text with information flows, and privacy washing can be read in these terms as a stated norm undermined by a documented flow within the same document. \citet{nissenbaum2011contextual} names the underlying structural tension the ``transparency paradox'': a notice simple enough to be understood cannot be complete, while one detailed enough to be accurate cannot be understood. The commitment-versus-fine-print structure we measure is a within-document instance of this tension, with the reassuring commitment playing the role of the simplified notice and the practice description the detailed one.

Legal doctrine provides parallel standards. Contract law's objective theory of interpretation, under which language is read as a reasonable person would understand it~\citep{farnsworth2004contracts}, and the FTC's ``net impression'' doctrine~\citep{ftc1983deception} both recognize that technically true statements can mislead when the overall impression is false. Privacy policies are often held not to be enforceable contracts; the analogy is to the interpretive standard, not to contractual enforceability, and the FTC deception framework, which does not require a contract, carries most of the doctrinal weight here. A pair qualifies as pragmatically contradictory when a reasonable consumer reading the commitment would form expectations that the documented practice violates.

In practice, the operational definition of contradiction in our pipeline is the judge prompt (Appendix~\ref{appendix:judge}), which implements the conceptual definition through six exclusion criteria: pairs are not contradictions when the practice implements or restates the commitment, is unrelated to it, concerns a different data type, user group, or context, realizes a possibility the commitment's hedging language permits, or concerns a different product or service. The within-document and directional properties are enforced structurally (only within-policy COMMITMENT--PRACTICE pairs are evaluated); the pragmatic property is operationalized by the prompt's criteria. These criteria track the theoretical grounding above. The hedging exclusion corresponds to the absence of a scalar implicature of exclusion: a commitment phrased with ``may'' implicates no upper bound for a practice to violate. The different-data-type, user-group, and context exclusions correspond to distinct informational norms under contextual integrity, under which no felicity failure arises. The qualifier-coverage filter (Section~\ref{sec:methodology}) corresponds to an implicature the commitment itself explicitly cancels. Every reported number reflects this operationalization, and Section~\ref{sec:results} notes cases where panel behavior diverged from the prompt's instructions.

\subsection{Observed Contradiction Patterns}

Empirical analysis of panel-confirmed contradictions across the judged OPPT companies reveals recurring patterns. We preview them here qualitatively to ground the definition in recurring category patterns; counts and category distributions appear in Section~\ref{sec:results}, produced by the pipeline detailed in Section~\ref{sec:methodology}. We derived these patterns through qualitative coding of pipeline results: we reviewed all panel-confirmed contradiction pairs, categorized each by the commitment and practice categories involved, and iteratively grouped cases into recurring structural patterns. Because the patterns are derived from pipeline output, they describe what the pipeline found; they are not an independent framework against which the pipeline can be evaluated.

The most prevalent patterns involve third-party sharing: THIRD\_PARTY commitments contradicted by THIRD\_PARTY practices, and SALE\_SHARING commitments (``we do not sell or share'') contradicted by THIRD\_PARTY practices. The SALE\_SHARING pattern is driven by CCPA-mandated disclosures: companies add ``we do not sell'' statements to comply with California requirements, but these blanket claims conflict with pre-existing descriptions of data sharing with affiliates, service providers, and advertising partners.

FIRST\_PARTY commitments contradicted by FIRST\_PARTY practices typically involve data collection scope: a commitment to collect ``only necessary information'' contradicted by extensive collection practices. Cross-category patterns, such as FIRST\_PARTY commitments contradicted by THIRD\_PARTY practices, capture cases where collection limitations are undermined by sharing practices.

A smaller set of patterns involves tracking and targeted advertising, such as commitments not to show behaviorally targeted advertising to children under 13 contradicted by advertising and attribution practices. No contradictions involving the SENSITIVE\_DATA category were panel-confirmed in either corpus; whether this null result reflects genuinely careful drafting around sensitive data, extraction behavior, or the category's definition warrants investigation (Section~\ref{sec:limitations}).

Qualitative examination reveals recurring structural patterns: multi-division policies where product-specific commitments conflict with enterprise-wide practices, and regulatory accretion where CCPA-mandated disclosures create new contradictions with pre-existing practice descriptions. Acquisition and transfer clauses that grant acquiring entities rights undermining prior commitments are a hypothesized third pattern; they generated many candidate pairs in preliminary experiments but are not represented among the confirmed contradictions of the enhanced pipeline (Section~\ref{sec:structural}).

\subsection{Relationship to Other Dark Patterns}

Privacy washing is distinct from dark patterns in user interfaces~\citep{mathur2019dark}, which manipulate behavior through design, and from jurisdiction-siloed disclosure~\citep{brackin2026jurisdiction}, which concerns where information is placed rather than what it says. These are complementary: a single policy can fragment information by jurisdiction while containing internal contradictions. Our broader class of pragmatic contradictions includes logical negation as a special case while capturing the reassurance-versus-practice tension that symbolic methods do not target.

\section{Methodology}
\label{sec:methodology}

\subsection{Pipeline Overview}

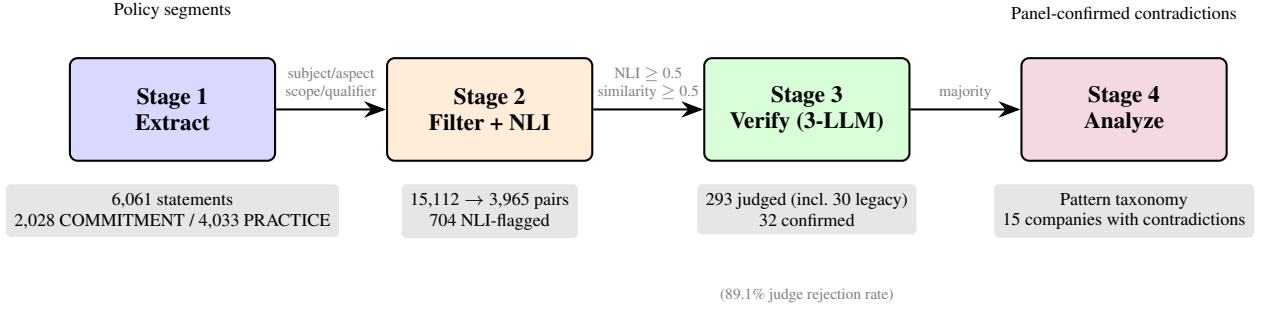
\begin{figure}[t]
\centering
\resizebox{\textwidth}{!}{%
\begin{tikzpicture}[
    node distance=0.8cm and 1.2cm,
    stage/.style={
        rectangle, rounded corners=3pt, draw=black, thick,
        minimum width=2.8cm, minimum height=1.4cm, align=center,
        font=\small\bfseries
    },
    count/.style={
        rectangle, rounded corners=2pt, fill=gray!20,
        minimum width=2.4cm, minimum height=0.6cm, align=center,
        font=\scriptsize
    },
    arrow/.style={-{Stealth[length=3mm]}, thick},
    filter/.style={font=\tiny, text=gray}
]

% Stage 1: Extract
\node[stage, fill=blue!15] (extract) {Stage 1\\Extract};
\node[count, below=0.3cm of extract] (c1) {6,061 statements\\2,028 COMMITMENT / 4,033 PRACTICE};

% Stage 2: Detect
\node[stage, fill=orange!15, right=1.5cm of extract] (detect) {Stage 2\\Filter + NLI};
\node[count, below=0.3cm of detect] (c2) {15,112 $\to$ 3,965 pairs\\704 NLI-flagged};

% Stage 3: Verify
\node[stage, fill=green!15, right=1.5cm of detect] (verify) {Stage 3\\Verify (3-LLM)};
\node[count, below=0.3cm of verify] (c3) {293 judged (incl.\ 30 legacy)\\32 confirmed};

% Stage 4: Analyze
\node[stage, fill=purple!15, right=1.5cm of verify] (analyze) {Stage 4\\Analyze};
\node[count, below=0.3cm of analyze] (c4) {Pattern taxonomy\\15 companies with contradictions};

% Arrows with filter labels
\draw[arrow] (extract) -- (detect) node[midway, above, filter, align=center] {subject/aspect\\scope/qualifier};
\draw[arrow] (detect) -- (verify) node[midway, above, filter, align=center] {NLI $\geq$ 0.5\\similarity $\geq$ 0.5};
\draw[arrow] (verify) -- (analyze) node[midway, above, filter] {majority};

% Judge rejection rate below verify counts
\node[filter, below=0.6cm of c3] {(89.1\% judge rejection rate)};

% Input/output labels
\node[above=0.4cm of extract, font=\scriptsize] {Policy segments};
\node[above=0.4cm of analyze, font=\scriptsize] {Panel-confirmed contradictions};

\end{tikzpicture}%
}
\caption{Four-stage pipeline architecture for privacy washing detection (OPPT corpus example counts shown). Category and self-segment filtering reduce 95,585 raw commitment-practice pairs to 53,797; metadata compatibility filtering (subject, aspect, scope, qualifiers) leaves 15,112; a semantic similarity pre-filter leaves 3,965 for NLI scoring. The 293 judged pairs include 30 legacy pairs carried over from an earlier pair-generation pass (Section~\ref{sec:methodology}). Multi-model judge verification identifies 32 panel-confirmed contradictions across 15 companies as run; the reproducible subset excluding legacy pairs retains 25 confirmations across 12 companies.}
\label{fig:pipeline}
\end{figure}

Our detection pipeline comprises four stages: statement extraction, compatibility filtering and NLI screening, multi-model judge verification, and thematic analysis. Figure~\ref{fig:pipeline} illustrates the architecture with counts at each stage. Three design principles guide the pipeline.

The first principle is decomposition before comparison. Comparing full policy segments (median 97 words for OPPT, interquartile range 47--214, with a long right tail reaching 5,545 words) produces noisy NLI signals because a single segment may contain both commitments and practices. Decomposing segments into atomic statements of approximately 10--30 words isolates individual commitment-practice pairs, producing more focused NLI input. In preliminary experiments, segment-level comparison generated systematic false positives: security implementations flagged against security commitments, restated commitments appearing contradictory, and informational content paired with active commitments. In an unquantified comparison, statement-level decomposition with typed pairing removed all three patterns from the cases we examined (the segment-level pilot artifacts are included in the released repository).

The second principle is typed pairing. Only COMMITMENT statements are paired with PRACTICE statements, eliminating false positives from complementary statements and focusing detection on the directional pattern that defines privacy washing.

The third principle is progressive filtering. Each pipeline stage reduces the candidate set: category filtering, metadata filtering, similarity scoring, NLI scoring, and multi-model judging progressively narrow candidates from tens of thousands of raw commitment-practice pairs to 956 sent to judges (293 OPPT + 663 OPP-115), yielding 111 panel-confirmed contradictions. This filtering matters because NLI alone is unreliable in both directions: it substantially over-flags non-contradictions, and pilot evidence indicates it also misses pairs judges confirm (Section~\ref{sec:pilot}). The judge panel addresses the former; the latter bounds the pipeline's recall. Throughout the paper we use one term per stage: \emph{generated pairs} (all COMMITMENT--PRACTICE pairs), \emph{compatibility-filtered pairs} (surviving category and metadata filters), \emph{similarity-filtered pairs} (surviving the semantic similarity pre-filter), \emph{NLI-flagged pairs} (NLI contradiction score $\geq$ 0.5), \emph{judged pairs} (sent to the judge panel), and \emph{panel-confirmed contradictions} (majority contradiction verdict).

One design caveat applies throughout: in the primary runs, the same three models serve as both extractors (Stage~1) and judges (Stage~3), so shared model biases could propagate across stages. Section~\ref{sec:limitations} discusses this concern in detail, and Section~\ref{sec:stability} reports a re-run with fully separated extraction and judge panels.

\subsection{Stage 1: Statement Extraction}

The first stage decomposes privacy policy segments into atomic statements using a three-LLM consensus panel comprising Claude Haiku 4.5 (Anthropic), GPT-5 mini (OpenAI), and Gemini 3 Flash Preview (Google). Each segment is processed independently by all three models in parallel, and statements are included only if extracted by at least two models. Cross-model statement matching uses cosine similarity on all-MiniLM-L6-v2 sentence embeddings~\citep{reimers2019sentencebert} (threshold 0.7), the same embedding model used in the Stage 2 similarity pre-filter, to align semantically equivalent extractions; this matching threshold was not varied in sensitivity analysis, so all downstream counts are conditional on it. This consensus approach reduces extraction noise and ensures that included statements reflect agreement across model architectures.
Preliminary single-model experiments revealed systematic false positive patterns (subject mismatches, aspect mismatches, and missing qualifiers) that motivated the structured metadata extraction described below.

The extraction schema captures structured metadata fields (the full extraction prompt appears in Appendix~\ref{appendix:extraction}):

\begin{itemize}
    \item \textbf{subject}: WHO performs the action: COMPANY, SERVICE\_PROVIDER, THIRD\_PARTY, AFFILIATES, or USER.
    \item \textbf{aspect}: WHAT data lifecycle stage: COLLECTION, USE, SHARING, SALE, RETENTION, DELETION, ACCESS\_CONTROL, or SECURITY.
    \item \textbf{scope}: Under WHAT conditions: UNIVERSAL, CONDITIONAL, CONSENT\_BASED, LEGAL\_REQUIREMENT, or GEOGRAPHIC\_LIMITED. Practices with LEGAL\_REQUIREMENT scope (e.g., ``we share when required by law'') are excluded from pairing, as they represent standard legal carve-outs rather than privacy washing.
    \item \textbf{qualifiers}: Array of limiting phrases verbatim from source text (e.g., ``except as required by law'', ``only for service delivery'').
\end{itemize}

The binary COMMITMENT/PRACTICE typing is a simplification; prior work has shown that user choice and control statements constitute a distinct category in privacy policies~\citep{mysoresathyendra2017identifying,bannihattikumar2020finding}. Section~\ref{sec:limitations} discusses how this distinction affects our results.

For the OPPT corpus, extraction prompts include per-segment annotation blocks from the companion dataset~\citep{brackin2026jurisdiction}. These annotations were generated through a three-LLM consensus process that achieved 91\% average agreement across models; 280 cases where models disagreed underwent human adjudication. The \texttt{Does/Does Not} annotation field (\texttt{does\_does\_not} in the released annotation files), which indicates whether a segment describes an action the company performs or refrains from, is included in these blocks as a typing hint. Its practical influence is limited: ``Does Not'' votes are rare in the attribute data (79 of 4,766 votes), and only 62 of the 2,028 OPPT COMMITMENT statements (3.1\%) originate from segments carrying any ``Does Not'' annotation. COMMITMENT/PRACTICE typing is therefore predominantly LLM-inferred for OPPT.

For OPP-115, we configured extraction to use the original human annotations' \texttt{Does/Does Not} polarity attribute to guide typing. A post-hoc audit conducted during revision found, however, that a segment-identifier mismatch between the attribute file and the segment file caused this annotation join to fail silently: no OPP-115 statement received annotation guidance, and all 4,975 statements were typed by the extraction models alone. Claims of annotation-guided typing therefore do not apply to the OPP-115 results reported here. Prompted by this failure, we audited the OPPT annotation join and confirmed it operated correctly: 6,048 of 6,061 OPPT statements carry annotation-block provenance, and every source segment identifier in the OPPT statement set resolves in the released attribute file. The low OPPT guidance rate reflects the rarity of ``Does Not'' votes in the source annotations, not a join failure.

Segments in the OTHER, POLICY\_CHANGE, and REGIONAL categories are excluded from extraction in both corpora, as they do not contain actionable data practice claims. Extraction produced 6,061 statements for OPPT (2,028 commitment, 4,033 practice) and 4,975 for OPP-115 (1,678 commitment, 3,297 practice). The consensus requirement and structured metadata criteria eliminate redundant and fragmentary statements.

\subsubsection{Category Mapping for OPP-115}

Our extraction categories were informed by OPP-115's annotation scheme but adapted to reflect contemporary regulatory frameworks. Six categories map directly: FIRST\_PARTY (First Party Collection/Use), THIRD\_PARTY (Third Party Sharing/Collection), USER\_ACCESS (User Access, Edit and Deletion), RETENTION (Data Retention), SECURITY (Data Security), and USER\_CHOICE (User Choice/Control).\footnote{The extraction models occasionally emit the alias labels DATA\_RETENTION and USER\_RIGHTS for the RETENTION and USER\_CHOICE categories; the pipeline treats each alias pair as equivalent, and all four labels are excluded from pairing (Section~\ref{sec:scope}).} Three categories, SALE\_SHARING, SENSITIVE\_DATA, and AUTOMATED\_DECISIONS, reflect post-2016 regulatory developments~\citep{ccpa2018,gdpr2016} absent from the original OPP-115 scheme. TRACKING subsumes OPP-115's narrower Do Not Track category. The remaining OPP-115 category, International and Specific Audiences, maps to INTL\_SPECIFIC and is excluded from pairing (Section~\ref{sec:scope}).

SALE\_SHARING reflects the CCPA's (2018) requirement that businesses disclose whether they ``sell'' personal information, a regulatory distinction absent when OPP-115 was created. When processing OPP-115, our extraction assigns SALE\_SHARING labels to sale-related statements; the original OPP-115 scheme captured such statements under Third Party Sharing/Collection without the specific ``sale'' framing. This temporal mismatch is a limitation when comparing results across corpora; however, longitudinal evidence indicates that the underlying data practices (disclosure to third parties for commercial purposes) were widespread in pre-CCPA policies, described in different terms~\citep{linden2020privacy,amos2021privacy}.

OPP-115 comprises 115 website privacy policies with 3,792 total segments across 10 categories. Each segment received fine-grained annotations, with each category containing multiple attributes (data type, collection mechanism, purpose), yielding over 23,000 individual attribute-value annotations~\citep{wilson2016creation}. Our extraction operates at segment-level category granularity rather than attribute-level annotation depth. Of the 3,792 total segments, we excluded 914 (24.1\%) from processing: 801 in the OTHER category (general statements not constituting data practice claims) and 113 in POLICY\_CHANGE (modification notifications); the REGIONAL category does not occur in this corpus. The retained 2,878 segments produced 4,975 atomic statements for analysis.

\subsection{Stage 2: Compatibility Filtering and NLI Screening}

The second stage pairs commitment statements with practice statements and scores each pair for contradiction using natural language inference. Structured metadata enables compatibility filtering that removes most candidate pairs before scoring; its effect on precision is untested (Section~\ref{sec:metadata}).

\subsubsection{Metadata-Based Filtering}

Before NLI scoring, we apply compatibility filters using the structured metadata:

\begin{itemize}
    \item \textbf{Subject compatibility}: Statements must have compatible subjects. COMPANY and AFFILIATES can pair (related entities), but COMPANY cannot pair with THIRD\_PARTY or SERVICE\_PROVIDER (different actors).
    \item \textbf{Aspect compatibility}: Statements must have related aspects. COLLECTION pairs with USE, SHARING, SALE, and RETENTION (data flow lifecycle), but ACCESS\_CONTROL does not pair with COLLECTION, and SECURITY does not pair with SHARING.
    \item \textbf{Scope filter}: Practices with LEGAL\_REQUIREMENT scope are excluded from pairing. A practice stating ``we share data when required by law'' represents a standard legal carve-out, not a contradiction of a privacy commitment.
    \item \textbf{Qualifier coverage}: If a commitment's qualifiers explicitly cover the practice's scope, the pair is excluded. For example, a commitment ``we do not share, except with service providers'' cannot be contradicted by a practice about service provider sharing.
\end{itemize}

Qualifier coverage uses exact string matching against qualifiers extracted verbatim during decomposition. The matching vocabulary comprises the 43 qualifier patterns observed at frequency $\geq 8$ in the extracted statements (e.g., ``required by law,'' ``with your consent,'' ``in connection with a merger''); the full pattern list and matching code are included in the released artifacts. This prioritizes precision over recall: paraphrased equivalents (``required by law'' vs.\ ``mandated by applicable regulations'') may be missed, but the downstream judge panel catches resulting false positives.

For OPPT, the pipeline generates 95,585 raw within-company commitment-practice pairs. Category and self-segment filtering leave 53,797; the four metadata compatibility filters (subject 24,324 pairs removed, aspect 12,758, scope 578, qualifier coverage 1,025, applied sequentially in that order) leave 15,112; and the semantic similarity pre-filter described below leaves 3,965. The metadata filters thus remove 71.9\% of category-compatible pairs, and the full cascade removes 95.9\% of raw pairs before NLI scoring. For OPP-115, 65,021 raw pairs reduce to 41,243 after category and self-segment filtering and to 9,877 after the similarity pre-filter; we did not enable the metadata compatibility filters for the OPP-115 run. The OPP-115 statements carry the same populated metadata fields (subject, aspect, scope, qualifiers) as OPPT statements, so a matched-configuration comparison is feasible and remains future work (Sections~\ref{sec:cross_temporal} and~\ref{sec:future}).

\subsubsection{Semantic Similarity and NLI Scoring}

Following filtering, a semantic similarity pre-filter further reduces noise. We compute embeddings using the all-MiniLM-L6-v2 sentence transformer~\citep{reimers2019sentencebert}. A single similarity threshold of 0.5 governs judge submission, applied uniformly to all pairs. The detection stage applies a lower pre-filter threshold of 0.3 to same-category pairs; because judge submission applies 0.5 uniformly, this lower threshold affects only which pairs receive NLI scores (and hence the NLI-flagged counts), not which pairs are judged or confirmed. We selected these thresholds from preliminary observations on the evaluation corpus without a held-out validation set, a circularity that Section~\ref{sec:sensitivity} partially addresses through a sensitivity analysis varying the judge-submission threshold; that analysis is one-sided, since no pair below 0.50 was ever judged. The similarity distribution of panel-confirmed contradictions reveals that 40.6\% cluster at the 0.50--0.55 boundary, indicating that threshold selection affects which cases reach evaluation.

For NLI scoring, we use the \texttt{cross-encoder/nli-deberta-v3-base} checkpoint,\footnote{\url{https://huggingface.co/cross-encoder/nli-deberta-v3-base}, a SentenceTransformers~\citep{reimers2019sentencebert} cross-encoder trained on SNLI and MultiNLI.} which fine-tunes the DeBERTaV3 architecture~\citep{he2023deberta} on SNLI and MultiNLI. Each pair is formatted as ``[CLS] commitment\_text [SEP] practice\_text [SEP]''. Pairs receiving a contradiction score of 0.5 or above are flagged as NLI-positive: 704 for OPPT and 1,781 for OPP-115. The 0.5 similarity threshold is then applied for judge submission. For OPP-115, the 663 judged pairs are exactly the NLI-flagged pairs meeting this threshold; the remaining 1,118 flagged pairs fell below it and were never judged. For OPPT, 263 of the 704 flagged pairs meet the threshold (441 fall below it).

The 293 judged OPPT pairs comprise these 263 plus 30 pairs carried over through the judge cache from an earlier pair-generation pass; the 30 legacy pairs are absent from the final pair set. Eight of the 33 pre-reclassification OPPT confirmations, including three of the four Google confirmations, come from these legacy pairs. We retain them because their verdicts are on statements extracted by the same pipeline, but flag the provenance for reproducibility. A post-hoc audit of the released artifacts explains the difference between the passes: all 30 legacy pairs satisfy the final configuration's similarity, NLI, subject, aspect, and scope criteria, and all 30 fail exactly one filter added after the earlier pass, the qualifier-coverage check and its standard-exception branch, which discards pairs whose practice statement matches consent-based, service-provider, or merger-and-acquisition carve-out patterns.

Legacy pairs confirm at 26.7\% (8/30) versus 9.5\% (25/263) for pairs in the final set: judges confirm, at nearly three times the base rate, precisely the class of pairs this filter discards as likely false positives. This discrepancy between filter design and judge behavior bears on the untested filter-precision question (Section~\ref{sec:metadata}). We retain the filter configuration as run rather than modifying it after observing judge verdicts, which would amount to post-hoc tuning against the judge panel; disabling the qualifier-coverage branch and judging the pairs it discards is a specific ablation we list in Section~\ref{sec:future}. Throughout the paper we report the as-run figures, 293 judged pairs with 32 panel-confirmed contradictions across 15 companies, as primary, since they are what the pipeline produced, and the reproducible subset, 263 judged pairs with 25 panel-confirmed contradictions (9.5\%) across 12 companies, as a robustness check (Section~\ref{sec:results}). (The single OPPT pair removed by the statement-type reclassification described below is itself a legacy pair.)

\subsection{Stage 3: Multi-Model Judge Verification}

The third stage provides the final precision layer. Even after metadata-based filtering, NLI alone over-flags pairs that involve topical tension without actual pragmatic contradiction.

To achieve precision, we employ a panel of three LLMs from different providers (Anthropic, OpenAI, Google; see the Data and Code Availability statement for model identifiers and experiment dates). All judges use temperature 0.0 for reproducibility. API requests were retried up to three times on transport errors with exponential backoff; a returned response whose verdict could not be parsed was recorded as invalid without re-request. Requests were routed through OpenRouter without pinning to a specific backend provider, a reproducibility caveat since provider-side serving changes can alter model outputs. Provider diversity reduces the risk of systematic bias but does not eliminate it, as models from different providers may share training data or exhibit correlated errors. All three judge providers also appear as companies in the OPPT corpus, so judge models evaluate statements from their own providers' policies; Section~\ref{sec:ethics} discusses this conflict of interest. Each judge receives a structured prompt asking whether the COMMITMENT and PRACTICE genuinely contradict each other, with guidance on distinguishing genuine contradictions from hedging language, scope differences, conditional exceptions, and product-boundary distinctions.

Majority vote determines the final verdict, following the self-consistency and multi-model panel-judging designs in prior work~\citep{wang2023selfconsistency,verga2024juries}: a pair is classified as a contradiction if at least two of three judges agree. For OPPT, 293 pairs were sent to judges with 83.6\% receiving unanimous verdicts and 16.4\% majority verdicts (Fleiss' $\kappa$~\citep{fleiss1971measuring} = 0.48). For OPP-115, 663 pairs were judged with 81.4\% unanimous and 18.4\% majority verdicts (Fleiss' $\kappa$ = 0.57). With three judges and a binary verdict, a majority exists whenever all three return valid responses; a single OPP-115 pair (0.2\%) lacked a usable majority because one judge returned an invalid response.

LLM judges exhibit systematic biases~\citep{zheng2023judging}, and judges from different providers may share such biases, so high inter-model agreement could reflect consistent bias patterns rather than accuracy against human judgment. These panel-confirmed findings serve as preliminary evidence; human expert validation with privacy law specialists is a necessary next step to establish ground truth precision.

Of the 293 OPPT pairs sent to judges, the panel initially confirmed 33; of the 663 OPP-115 pairs, 99. A statement-type reclassification audit (Section~\ref{sec:limitations}) removed confirmations involving statements retyped as user-capability rather than company-commitment, yielding 32 and 79. The removals are unevenly distributed: 1 of 33 OPPT confirmations (3.0\%) versus 20 of 99 OPP-115 confirmations (20.2\%). This asymmetry is consistent with the OPP-115 run's lack of metadata compatibility filtering and is further evidence that the two runs are not configuration-matched (Section~\ref{sec:cross_temporal}).

The confirmation rates used throughout (10.9\% and 11.9\%) divide the corrected numerators by the uncorrected denominators. Because the audit was applied only to confirmed pairs, some rejected pairs may also involve retyped statements, so these rates are lower bounds on the fully corrected confirmation rate (the uncorrected rates are 11.3\% and 14.9\%).

As a sanity check, the author reviewed all 111 panel-confirmed contradictions, examining each pair's extracted statements, judge reasoning, and source policy context. This review found no cases that appeared clearly erroneous, including those involving regulatory exceptions like merger-and-acquisition (M\&A) clauses or disclosures permitted under the Health Insurance Portability and Accountability Act (HIPAA) (Section~\ref{sec:limitations}). This single-reviewer, non-blind assessment does not constitute rigorous validation; it serves only as a preliminary quality check. Independent validation by multiple privacy law experts using blind methodology would be necessary to establish ground-truth precision. This precision-oriented check also coexists with the pilot's documented judge failure modes (Section~\ref{sec:pilot}) without conflict: those failure modes are recall failures (judges dismissing pairs a human reviewer classified as genuine), whereas this review probes precision among confirmed pairs. The one precision-relevant pattern it did not flag as clearly erroneous is the product-boundary confirmations discussed in Section~\ref{sec:results}.

\subsection{Computational Requirements and Scalability}
\label{sec:computational}

For the OPPT corpus (123 companies, 3,651 segments, 293 pairs sent to judges), the complete pipeline required approximately 2 hours of wall-clock processing at a total API cost of approximately \$8 across three providers (January 2026 prices). Category, metadata, and similarity filtering together remove 95.9\% of raw pairs before expensive judge verification.

This filtering eliminates the majority of false positive candidates at near-zero marginal cost, so adding new policies increases extraction cost linearly while judge verification cost grows only with the small number of genuinely contradictory pairs.

The current system is suitable for regulatory audits of specific companies or sectors. At the measured per-company cost (roughly \$0.07), processing a corpus the size of the published PrivaSeer corpus of over one million policies~\citep{srinath2021corpus} would cost on the order of \$70,000 in API fees, dominated by extraction; this estimate excludes the cost of producing the segment-level annotations that guide statement typing, which the PrivaSeer corpus does not include, and the practical barriers are corpus access and validation rather than compute. Larger-scale deployment would also benefit from caching frequently occurring qualifier patterns and pre-computing subject/aspect compatibility matrices.

\subsection{Detection Scope and Exclusions}
\label{sec:scope}

The pipeline is designed to detect pragmatic contradictions between commitments and practices within single policy documents and does not attempt comprehensive contradiction detection across all categories or types. Several classes of contradiction fall outside scope.

The pipeline restricts pairing to six actionable categories: FIRST\_PARTY, THIRD\_PARTY, TRACKING, SALE\_SHARING, SENSITIVE\_DATA, and AUTOMATED\_DECISIONS. All other categories are excluded from pairing: principally SECURITY, RETENTION, USER\_CHOICE, and USER\_ACCESS (together with their alias labels, Section~\ref{sec:methodology}), along with low-frequency categories such as INTL\_SPECIFIC (100 OPPT statements, 83 OPP-115). The excluded categories primarily describe implementation mechanisms (encryption methods, deletion timelines, rights-exercise procedures) rather than substantive data flows. This filtering eliminates false positives from complementary procedure-and-implementation pairs, though it may exclude some meaningful contradictions (e.g., retention commitments contradicted by archival practices, or security commitments contradicted by unencrypted storage).

Two of the six actionable categories proved inert in practice: SENSITIVE\_DATA (67 OPPT statements; 28 judged pairs, 0 confirmed) and AUTOMATED\_DECISIONS (40 OPPT statements; 1 judged pair, 0 confirmed), and the OPP-115 extraction emitted no statements in either category, so both are effectively OPPT-only. We discuss the SENSITIVE\_DATA null result in Section~\ref{sec:limitations}.

In preliminary experiments, including the excluded categories generated over 1,200 additional candidate pairs across both corpora, of which 47 reached the judge panel; none were confirmed as contradictions. The 95\% Clopper--Pearson interval~\citep{clopper1934use} for 0/47 is 0--7.5\%, which is below the 11\% rate for included categories but does not exclude a modest true rate. The exclusion is therefore a precision-motivated design choice rather than definitive evidence that these categories contain no contradictions.

The pipeline targets pragmatic contradictions (reassurance undermined by practice), not logical negations requiring formal ontologies. Contradictions of the form ``we do not collect email'' versus ``we collect email'' fall within the scope of symbolic approaches like PolicyLint~\citep{andow2019policylint}; our approach captures the more prevalent pattern of ``we minimize collection'' versus ``we collect [extensive list]'' that pragmatic reasoning distinguishes but symbolic methods miss.

The pipeline detects internal policy contradictions only, not code-versus-policy consistency~\citep{andow2020policheck,zimmeck2019maps}, policy-versus-regulation compliance or completeness~\citep{ikhwantri2025exclaim,torre2020aiassisted,amaral2022aienabled}, or cross-document contradictions.

Finally, the pipeline's scope excludes the user-experience dimension of contradiction (Section~\ref{sec:limitations}).

\section{Datasets}
\label{sec:dataset}

\subsection{The OPPT Corpus: Origin and Composition}

The OPPT corpus was collected for our companion work on jurisdiction-siloed disclosure~\citep{brackin2026jurisdiction}. It comprises privacy policies from 123 companies collected January 11--13, 2026, a three-day snapshot window, more than three years after both the Sephora CCPA settlement (August 2022) and the California Privacy Rights Act (CPRA) effective date (January 1, 2023)~\citep{cpra2020}. This three-year interval postdates the post-Sephora and post-CPRA compliance deadlines, making detected contradictions less attributable to transitional compliance states. The corpus spans diverse industries including technology platforms, financial services, healthcare, retail, travel, and data brokers. Table~\ref{tab:corpus} summarizes corpus composition after statement extraction.

\begin{table}[htbp]
\centering
\begin{tabular}{lrrrr}
\toprule
\textbf{Company} & \textbf{Segments} & \textbf{Statements} & \textbf{Commitments} & \textbf{Practices} \\
\midrule
Meta & 191 & 321 & 32 & 289 \\
Microsoft & 66 & 239 & 68 & 171 \\
Khan Academy & 64 & 147 & 56 & 91 \\
Hilton & 81 & 141 & 69 & 72 \\
American Airlines & 60 & 138 & 55 & 83 \\
\emph{Other 118 companies} & 3,189 & 5,075 & 1,748 & 3,327 \\
\midrule
\textbf{Total (123 companies)} & 3,651 & 6,061 & 2,028 & 4,033 \\
\bottomrule
\end{tabular}
\caption{OPPT corpus composition after statement extraction. Top 5 companies by statement count shown.}
\label{tab:corpus}
\end{table}

\subsection{Annotation Guidance and Its Limits}

Both corpora were processed with extraction prompts designed to incorporate per-segment annotations. In practice, this guidance was minimal for OPPT (only 3.1\% of COMMITMENT statements originate from ``Does Not''-annotated segments) and inoperative for OPP-115, where a segment-identifier mismatch caused the annotation join to fail (Section~\ref{sec:methodology}). OPP-115's \texttt{Does/Does Not} polarity attribute (692 ``Does Not'' annotations across 219 segments, 4.9\% of annotations; figures computed from the released corpus annotations) remains a promising typing signal for future work. The reliability of the attribute itself is uncertain: \citet{wilson2016creation} report segment-level Fleiss' $\kappa$ per category ranging from 0.49 to 0.91 (their Table~2) but no agreement figure for the \texttt{Does/Does Not} attribute specifically, and state (their Section~4.2) that during consolidation attribute values were merged by majority vote and set to Unspecified in approximately a third of all mergers, an indication of substantial attribute-level disagreement.

\subsection{Company Selection Rationale}

The 123 companies were selected for the companion jurisdiction-siloed disclosure study~\citep{brackin2026jurisdiction} based on diversity of policy styles, organizational structures, and industry sectors. The corpus spans social media (Meta, LinkedIn, TikTok), telecommunications (T-Mobile, AT\&T, Verizon), financial services (Venmo, PayPal, Equifax), healthcare, travel (Hilton, Delta), retail (Walmart, Target), data brokers (Clearview AI, Gravy Analytics), and AI companies (OpenAI, Anthropic, Perplexity). Policy sizes range from 7 segments to 191 segments (Meta), with a median of 24 segments per company. The contradiction analysis was conducted post-hoc on this existing corpus. Because companies were purposively selected for policy diversity rather than randomly sampled, the prevalence estimates in this paper describe these corpora and should not be generalized to any broader population of companies.

\section{Results}
\label{sec:results}

\subsection{Pipeline Performance}

Table~\ref{tab:pipeline} summarizes the pipeline's filtering efficiency. Preliminary experiments with single-model extraction and minimal metadata produced high false positive rates; iterative refinement with three-LLM consensus extraction and structured metadata filtering reduced the candidate set reaching judges by orders of magnitude, though its effect on precision is untested (Section~\ref{sec:metadata}).

\begin{table}[htbp]
\centering
\begin{tabular}{lrr}
\toprule
\textbf{Stage} & \textbf{OPPT} & \textbf{OPP-115} \\
\midrule
Statements extracted & 6,061 & 4,975 \\
Generated pairs & 95,585 & 65,021 \\
After category and self-segment filtering & 53,797 & 41,243 \\
Compatibility-filtered pairs (metadata) & 15,112 & --- \\
Similarity-filtered pairs & 3,965 & 9,877 \\
NLI-flagged pairs & 704 & 1,781 \\
Judged pairs (as run) & 293 & 663 \\
\quad excluding legacy pairs & 263 & --- \\
Panel-confirmed contradictions & 32 & 79 \\
\quad excluding legacy pairs & 25 & --- \\
Judge confirmation rate & 10.9\% & 11.9\% \\
\quad excluding legacy pairs & 9.5\% & --- \\
\bottomrule
\end{tabular}
\caption{Pipeline filtering stages for both corpora. OPPT applies the metadata compatibility filters (subject, aspect, scope, qualifiers); the OPP-115 run was configured without them, so its similarity pre-filter operates directly on the category-filtered pairs. Multi-model judge verification provides the final precision layer for both corpora. Primary OPPT rows report the pipeline as run; the ``excluding legacy pairs'' rows report the subset reproducible from the released final pair set, which omits 30 pairs carried over from an earlier pair-generation pass (Section~\ref{sec:methodology}). OPP-115 has no legacy pairs.}
\label{tab:pipeline}
\end{table}

Some panel-confirmed contradictions involve regulatory carve-outs (e.g., M\&A transfer clauses, HIPAA-permitted disclosures); Section~\ref{sec:limitations} explains why these remain findings.

\subsection{Per-Company Findings}

Two terms recur in what follows: we use \emph{full-corpus rate} for the share of all corpus companies with at least one panel-confirmed contradiction and \emph{judged-company rate} for the share among companies with judged pairs. As run, 15 of 123 OPPT companies contain at least one panel-confirmed contradiction (12.2\%; 95\% Clopper--Pearson CI: 7.0--19.3\%), a judged-company rate of 15 of 60 (25\%). In the reproducible subset that excludes legacy pairs (Section~\ref{sec:methodology}), the full-corpus rate is 9.8\% (12/123; 95\% CI: 5.1--16.4\%); Bumble, Khan Academy, and Tesla each had a single confirmation that came from a legacy pair.

All intervals in this paper describe binomial variability under hypothetical resampling of the same corpora; they are not inferences about any broader population of companies, since the corpora were purposively constructed (Section~\ref{sec:dataset}). They also capture sampling variation only, not classification error, which is unquantified and plausibly dominant given that precision against human judgment is unknown.

Of the 63 unjudged companies, 45 produced no NLI-flagged pairs, while 18 produced NLI-flagged pairs that all fell below the 0.5 similarity threshold applied at judge submission and were therefore never evaluated. The full-corpus figure is a lower bound: 18 of the unjudged companies were excluded by a threshold rather than by absence of signal, the 45 zero-flag companies may contain contradictions the filtering stages did not surface, and the pipeline's aggressive filtering may miss contradictions expressed through substantially different vocabulary or requiring multi-hop inference. The judged-company rate, by contrast, is computed over a subset selected for having judged pairs and may overstate full-corpus prevalence.

We discuss the OPPT findings company by company, selecting illustrative examples for their distinct pattern types rather than by contradiction count. Raw counts should be read against the number of pairs each company had judged, which varies with policy length: Google's 4 confirmations came from only 4 judged pairs, whereas Khan Academy's single confirmation came from 34. Duolingo, Microsoft, and Roblox lead with 4 panel-confirmed contradictions each; Google also reaches 4, but three of its four come from legacy pairs, only one survives in the reproducible subset, and its verdicts carry the conflict-of-interest caveat discussed below. Microsoft's contradictions center on de-identification claims conflicting with identifier practices:

\begin{quote}
\small
\textbf{COMMITMENT:} Microsoft uses personal data in the least identifiable form necessary and relies on statistical and aggregated pseudonymized data for business operations.

\textbf{PRACTICE:} Microsoft transmits Tailored experiences data to Microsoft servers and stores it with unique identifiers to recognize individual users and understand device patterns.

\textbf{Verdict:} UNANIMOUS (similarity=0.51, NLI=1.00)
\end{quote}

Notion's single confirmed contradiction, unanimous across all three judges, illustrates the third-party advertising pattern with matched scope on both sides:

\begin{quote}
\small
\textbf{COMMITMENT:} The company does not disclose user information to advertise third-party products or services via the Services.

\textbf{PRACTICE:} The company discloses online identifiers to third parties through online advertising services in a manner that may constitute a sale or sharing under CCPA.

\textbf{Verdict:} UNANIMOUS (similarity=0.56, NLI=1.00)
\end{quote}

Walmart's contradictions illustrate a set-inclusion form, in which a universally quantified practice necessarily covers a specific carve-out:

\begin{quote}
\small
\textbf{COMMITMENT:} The company does not disclose text messaging opt-in and consent data to third parties for their marketing purposes.

\textbf{PRACTICE:} The company discloses all categories of personal information with certain categories of third parties.

\textbf{Verdict:} UNANIMOUS (similarity=0.54, NLI=1.00)
\end{quote}

All four pairs sent to judges for Google were confirmed as contradictions, though this 100\% rate reflects a small sample, three of the four are legacy pairs (only one survives in the reproducible subset), and because one judge model is produced by Google, these verdicts are also subject to the conflict of interest discussed in Section~\ref{sec:ethics}. All involve third-party sharing claims contradicted by partner integration practices. Roblox also produces four contradictions related to compliance with the Children's Online Privacy Protection Act (COPPA) and advertising disclosures.

Duolingo's four contradictions involve children's data and speech processing, and Uber's three involve Guest User marketing exclusions. Confirmed pairs from both companies include a product-boundary or conditional-scope form: a commitment scoped to a specific product or user group (Duolingo ABC, Guest Users) paired with a general practice statement that carries no corresponding carve-out. The judge prompt (Appendix~\ref{appendix:judge}) instructs judges that statements about different products or services are not contradictions; the judges instead treated the general practice statement as encompassing the scoped commitment, since the policy text leaves the scope relationship unspecified. Confirmations of this form show the panel applying the prompt's product-boundary instruction inconsistently, a discrepancy between the operational definition and panel behavior discussed in Section~\ref{sec:limitations}. We have not systematically counted how many of the 111 panel-confirmed contradictions take this form; quantifying it requires re-annotating each pair for product scope and is part of the planned expert validation (Section~\ref{sec:future}), and prevalence figures should be read with the possibility of such cases in mind.

Table~\ref{tab:oppt_companies} summarizes all 15 OPPT companies with panel-confirmed contradictions.

\begin{table}[htbp]
\centering
\begin{tabular}{lrrl}
\toprule
\textbf{Company} & \textbf{Count} & \textbf{Unanimous} & \textbf{Pattern} \\
\midrule
Duolingo & 4 & 3 & FIRST\_PARTY $\to$ THIRD\_PARTY \\
Microsoft & 4 & 2 & FIRST\_PARTY $\to$ FIRST\_PARTY \\
Google\textsuperscript{\dag} & 4 & 1 & THIRD\_PARTY $\to$ THIRD\_PARTY \\
Roblox & 4 & 1 & Mixed (no dominant pattern) \\
Uber & 3 & 2 & FIRST\_PARTY $\to$ TRACKING \\
GitHub & 2 & 1 & SALE\_SHARING $\to$ THIRD\_PARTY \\
Walmart & 2 & 1 & THIRD\_PARTY $\to$ THIRD\_PARTY \\
Appriss\textsuperscript{\dag} & 2 & 0 & SALE\_SHARING $\to$ THIRD\_PARTY \\
Notion & 1 & 1 & THIRD\_PARTY $\to$ SALE\_SHARING \\
Bumble\textsuperscript{\dag} & 1 & 0 & THIRD\_PARTY $\to$ THIRD\_PARTY \\
Khan Academy\textsuperscript{\dag} & 1 & 0 & SALE\_SHARING $\to$ THIRD\_PARTY \\
Meta & 1 & 0 & FIRST\_PARTY $\to$ FIRST\_PARTY \\
Motorola Solutions & 1 & 0 & SALE\_SHARING $\to$ SALE\_SHARING \\
Tesla\textsuperscript{\dag} & 1 & 0 & SALE\_SHARING $\to$ SALE\_SHARING \\
Venmo & 1 & 0 & SALE\_SHARING $\to$ THIRD\_PARTY \\
\midrule
\textbf{Total} & \textbf{32} & \textbf{12} & \\
\bottomrule
\end{tabular}
\caption{OPPT companies with unvalidated, LLM-panel-confirmed candidate contradictions; precision against human expert judgment has not been established (Section~\ref{sec:limitations}). ``Count'' = total panel-confirmed contradictions; ``Unanimous'' = unanimous (3/3) verdicts. For companies with a single confirmed contradiction, the ``Pattern'' column reports that case's category pair; for companies with multiple contradictions it reports the modal category pair, so entries do not sum to Table~\ref{tab:categories}. Of the 32 panel-confirmed contradictions, 12 (37.5\%) are unanimous and 20 (62.5\%) are majority decisions. \textsuperscript{\dag}Counts depend on legacy pairs (Section~\ref{sec:methodology}): in the reproducible subset, which retains 25 confirmations across 12 companies, Google and Appriss drop to 1 confirmation each, and Bumble, Khan Academy, and Tesla drop to 0.}
\label{tab:oppt_companies}
\end{table}

\subsection{Contradiction Pattern Analysis}

The category distribution of panel-confirmed contradictions concentrates in third-party sharing patterns, a concentration that largely reflects the composition of judged pairs (below). Table~\ref{tab:categories} presents the distribution of commitment-practice category pairs for OPPT.

\begin{table}[htbp]
\centering
\begin{tabular}{lrr}
\toprule
\textbf{Commitment Category $\to$ Practice Category} & \textbf{Count} & \textbf{\%} \\
\midrule
THIRD\_PARTY $\to$ THIRD\_PARTY & 6 & 18.8\% \\
FIRST\_PARTY $\to$ FIRST\_PARTY & 6 & 18.8\% \\
SALE\_SHARING $\to$ THIRD\_PARTY & 6 & 18.8\% \\
FIRST\_PARTY $\to$ THIRD\_PARTY & 3 & 9.4\% \\
TRACKING $\to$ FIRST\_PARTY & 2 & 6.2\% \\
FIRST\_PARTY $\to$ TRACKING & 2 & 6.2\% \\
SALE\_SHARING $\to$ SALE\_SHARING & 2 & 6.2\% \\
Other & 5 & 15.6\% \\
\bottomrule
\end{tabular}
\caption{Category pair distribution of panel-confirmed OPPT contradictions. Pairs in which both sides are THIRD\_PARTY or SALE\_SHARING categories (the THIRD\_PARTY $\to$ THIRD\_PARTY, SALE\_SHARING $\to$ THIRD\_PARTY, and SALE\_SHARING $\to$ SALE\_SHARING rows plus one THIRD\_PARTY $\to$ SALE\_SHARING case within Other) account for 15 of 32 findings (46.9\%); including pairs with third-party sharing on one side only raises this to 18 of 32 (56.3\%).}
\label{tab:categories}
\end{table}

Three category pairs tie as the most frequent, at 6 of 32 (18.8\%) each: THIRD\_PARTY $\to$ THIRD\_PARTY, FIRST\_PARTY $\to$ FIRST\_PARTY, and SALE\_SHARING $\to$ THIRD\_PARTY; no single pair dominates at this sample size. Aggregating by topic rather than by category pair, contradictions with third-party sharing on at least one side account for 18 of 32 confirmed OPPT cases (56.3\%) and 55 of 79 OPP-115 cases (69.6\%).

This concentration must be read against base rates, since FIRST\_PARTY and THIRD\_PARTY are also the largest categories among judged pairs. Using pre-reclassification counts, OPPT pairs with THIRD\_PARTY or SALE\_SHARING on either side were confirmed at 13.2\% (23/174) versus 8.4\% (10/119) for pairs involving neither category; for OPP-115 the rates are 15.7\% (65/413) versus 13.6\% (34/250). Neither lift is statistically significant (two-sided Fisher's exact tests, $p = 0.26$ for OPPT and $p = 0.50$ for OPP-115), so these samples cannot resolve whether judges confirm third-party sharing pairs at a genuinely higher rate. All significance tests in this paper are exploratory and unadjusted for multiple comparisons. Most of the category's dominance among confirmed contradictions instead reflects its share of judged pairs (a selection effect of the upstream NLI flagging and similarity filtering) rather than a judge-stage effect. The structural interpretation (tension between user-facing commitments to limit sharing and data-driven revenue models) is consistent with this pattern but not established by it. The upstream origin of the concentration also bears on the concern that the judge prompt's example, which resembles the SALE\_SHARING pattern, primed judges toward this category; Section~\ref{sec:limitations} discusses this further.

Panel-confirmed contradictions concentrate at lower similarity values in absolute terms (40.6\% fall in the 0.50--0.55 range), but so do judged pairs (42.7\% fall in the same range): the confirmation \emph{rate} by similarity bin is flat to noisy for OPPT (10.4\% at 0.50--0.55, 13.9\% at 0.55--0.60, 7.3\% at 0.60--0.65, with single-digit cell counts above 0.70). The concentration of confirmations at low similarity is therefore a base-rate property of the judged sample, not evidence that low-similarity pairs are more likely to be contradictions.

\subsection{Judge Agreement Analysis}

As reported in Section~\ref{sec:methodology}, the panel achieves high usable consensus; Table~\ref{tab:agreement} summarizes the agreement metrics. The three judges differ systematically in leniency: the Google judge votes CONTRADICTION on 6.5\% of OPPT and 11.6\% of OPP-115 pairs, roughly half the rates of the Anthropic (13.0\%, 19.6\%) and OpenAI (16.4\%, 20.5\%) judges. Pairwise Cohen's $\kappa$~\citep{cohen1960coefficient} values range from 0.40 to 0.59, with the two Google pairings lowest on OPPT. This conservatism propagates to the confirmations: 16 of the 33 pre-reclassification OPPT confirmations (48\%) were 2--1 decisions over the Google judge's dissent, so nearly half of all confirmations rest on the other two judges outvoting the most conservative one; panel composition, not only the underlying phenomenon, shapes the reported counts.

Of the 32 panel-confirmed OPPT contradictions, 12 (37.5\%) are unanimous and 20 (62.5\%) are majority decisions. The higher majority-vote proportion among panel-confirmed contradictions compared to rejections is consistent with contradictions requiring contextual reasoning on which models differ, though this asymmetry is also what low prevalence predicts. Under a stricter unanimous-only criterion, 12 OPPT contradictions across 8 companies (8 of 60 judged companies, 13\%) and 49 OPP-115 contradictions across 30 companies (30 of 82 judged companies, 37\%) would remain. The range from the unanimous-only to the majority criterion (12 to 32 OPPT confirmations, 8 to 15 companies) is the uncertainty band attributable to the aggregation rule alone, and it is wider than the binomial intervals reported in Section~\ref{sec:results}.

\begin{table}[htbp]
\centering
\begin{tabular}{lrr}
\toprule
\textbf{Metric} & \textbf{OPPT} & \textbf{OPP-115} \\
\midrule
Pairs judged & 293 & 663 \\
Unanimous (3/3) & 83.6\% (245) & 81.4\% (540) \\
Majority (2/3) & 16.4\% (48) & 18.4\% (122) \\
Split/insufficient & 0\% (0) & 0.2\% (1) \\
Usable consensus & 100\% & 99.8\% \\
Fleiss' $\kappa$ & 0.48 & 0.57 \\
PABAK & 0.78 & 0.75 \\
\midrule
CONTRADICTION rate, Anthropic judge & 13.0\% & 19.6\% \\
CONTRADICTION rate, OpenAI judge & 16.4\% & 20.5\% \\
CONTRADICTION rate, Google judge & 6.5\% & 11.6\% \\
Cohen's $\kappa$, Anthropic--OpenAI & 0.59 & 0.55 \\
Cohen's $\kappa$, Anthropic--Google & 0.40 & 0.58 \\
Cohen's $\kappa$, OpenAI--Google & 0.42 & 0.58 \\
\bottomrule
\end{tabular}
\caption{Inter-rater agreement metrics for the 3-LLM judge panel, with per-judge CONTRADICTION verdict rates and pairwise Cohen's $\kappa$. PABAK is $2P_o - 1$, where $P_o$ is the mean pairwise observed agreement across the three judge pairings; see text.}
\label{tab:agreement}
\end{table}

The Fleiss' $\kappa$ values of 0.48 (OPPT) and 0.57 (OPP-115) are deflated by the low base rate ($\sim$11\%) of positive labels, which inflates expected chance agreement~\citep{feinstein1990high,byrt1993bias}. For reference, we also report the prevalence-adjusted bias-adjusted kappa (PABAK)~\citep{byrt1993bias}, the quantity $2P_o - 1$ where $P_o$ is the mean pairwise observed agreement across the three judge pairings (0.78 for OPPT, 0.75 for OPP-115); PABAK is defined for two raters, so this three-rater extension is descriptive rather than a validated statistic. Fleiss' $\kappa$ also assumes independent annotators, and LLM judges may share biases that render the framework inapplicable~\citep{pavlick2019inherent}. The observed agreement and unanimous verdict rates (83.6\%, 81.4\%) are therefore the more interpretable consensus metrics.

Qualitative examination of judge reasoning illustrates judges rejecting false positives that NLI cannot distinguish. For example, one pair with NLI score 0.9997 (Roblox: ``video data is not shared'' vs.\ ``we share username and email with event hosts'') was unanimously rejected as NOT\_CONTRADICTION by all three judges, each citing ``different data types'' in their reasoning. NLI detected topical tension (both involve sharing) but could not distinguish video processing from account identifier sharing. Across both corpora, 562 pairs with NLI scores above 0.95 (178 in OPPT, 384 in OPP-115) were unanimously rejected by judges, illustrating the role of multi-model verification as the pipeline's precision layer.

Two additional examples illustrate distinct false positive patterns. An American Airlines pair (NLI=0.9999) committed to retaining ``personal information only as long as reasonably necessary'' while practicing ``shredding or incinerating paper documents''; judges unanimously recognized these as complementary (retention duration vs.\ disposal method) rather than contradictory. A Kaleida Health pair (NLI=0.9998) committed to restricting ``protected health information'' (PHI) disclosure while permitting ``completely de-identified health information'' disclosure without restriction; judges reasoned that de-identified information falls outside the PHI commitment's scope under HIPAA definitions.

\subsection{Pilot Evidence on Judge and NLI Recall}
\label{sec:pilot}

The precision benefit of judging has a documented recall cost. In a pilot on an earlier, segment-level version of the pipeline (17 OPPT companies with 318 pairs and 43 OPP-115 companies with 820 pairs, superseded by the statement-level pipeline reported here), a single-reviewer manual assessment classified 45 of 71 NLI-flagged pairs as genuine contradictions, 19 as borderline, and 7 as false positives. The judge panel, using the same three models and similar exclusion criteria, dismissed 57 of the 71 flagged pairs, and for nine companies dismissed every pair the reviewer had classified as genuine (including 4 of 4 for The New York Times and 9 of 9 for Tesla). The Jaccard overlap between NLI-flagged and judge-confirmed sets was 7.7\%. Three judge failure modes recurred: narrowly literal readings of the commitment, rationalizing practices as consistent with broad claims, and dismissing pairs on a personally identifiable information (PII) versus non-PII distinction the commitment did not draw.

The complementary direction is worse for NLI: in the same pilot the panel confirmed 126 pairs, of which 112 (89\%) had NLI contradiction scores below the flagging threshold, many exactly 0.0.\footnote{The OPP-115 half of this tabulation reproduces exactly from the released pilot artifacts; the OPPT half derives from an intermediate judge file not retained, and recomputation on the closest surviving artifact yields 132 confirmed with 90\% unflagged, the same direction and magnitude.} NLI screening therefore functions as a severe recall bottleneck as well as a noisy precision signal.

Because the judge stage is unchanged in models and near-unchanged in prompt criteria, these dismissal patterns plausibly persist in the reported results (the assessment has not been re-run on the statement-level pipeline), which would mean the panel-confirmed counts understate true prevalence beyond the filtering-based lower-bound arguments made above. This cuts against treating the judge panel as approximating ground truth and is a further reason expert validation (Section~\ref{sec:limitations}) must precede any precision or recall claim.

\subsection{Threshold Sensitivity}
\label{sec:sensitivity}

The detection stage uses differentiated semantic similarity thresholds for candidate generation (0.5 for cross-category pairs, 0.3 for same-category pairs), while judge submission applies a uniform threshold of 0.5 (Section~\ref{sec:methodology}); the NLI contradiction threshold is 0.5. Because all scores are pre-computed for every judged pair, we can assess sensitivity by raising the uniform judge-submission threshold and measuring the effect on judged pairs and confirmed contradictions (Table~\ref{tab:sensitivity}). All counts reflect the post-reclassification totals (32 OPPT, 79 OPP-115) used throughout the paper.

\begin{table}[htbp]
\centering
\begin{tabular}{lrrrrr}
\toprule
\textbf{Sim.\ threshold} & \textbf{Judged} & \textbf{Confirmed} & \textbf{Rate} & \textbf{Companies} & \textbf{Lost} \\
\midrule
\multicolumn{6}{l}{\emph{OPPT}} \\
\textbf{0.50 (current)} & \textbf{293} & \textbf{32} & \textbf{10.9\%} & \textbf{15} & \textbf{---} \\
0.55 & 168 & 19 & 11.3\% & 12 & 13 \\
0.60 & 89 & 8 & 9.0\% & 6 & 24 \\
0.65 & 34 & 4 & 11.8\% & 4 & 28 \\
0.70 & 11 & 2 & 18.2\% & 2 & 30 \\
\midrule
\multicolumn{6}{l}{\emph{OPP-115}} \\
\textbf{0.50 (current)} & \textbf{663} & \textbf{79} & \textbf{11.9\%} & \textbf{42} & \textbf{---} \\
0.55 & 394 & 53 & 13.5\% & 28 & 26 \\
0.60 & 206 & 35 & 17.0\% & 22 & 44 \\
0.65 & 98 & 19 & 19.4\% & 11 & 60 \\
0.70 & 41 & 8 & 19.5\% & 4 & 71 \\
\bottomrule
\end{tabular}
\caption{Similarity threshold sensitivity analysis (uniform judge-submission threshold, post-reclassification counts). ``Judged'' = pairs reaching the judge panel; ``Confirmed'' = panel-confirmed contradictions; ``Lost'' = confirmed contradictions excluded relative to the current threshold. The OPPT confirmation rate is roughly stable across thresholds (high-threshold cells are small), while the OPP-115 rate rises monotonically from 11.9\% to 19.5\%, consistent with the bin-rate analysis in Section~\ref{sec:cross_temporal}. No pair below 0.50 was judged, so the analysis is one-sided and cannot speak to recall below the current threshold.}
\label{tab:sensitivity}
\end{table}

The similarity threshold exhibits an asymmetric sensitivity pattern. Raising the threshold above 0.50 rapidly excludes confirmed contradictions: at 0.55, 40.6\% of OPPT and 32.9\% of OPP-115 confirmations are lost; at 0.60, the losses reach 75.0\% and 55.7\% respectively. These losses follow from the concentration of judged pairs, and hence of confirmations, at lower similarity values. Within the primary runs, no pair below 0.50 was ever judged, so this sensitivity analysis cannot speak to recall in that region. The stability run (Section~\ref{sec:stability}) judges the full flagged set and finds confirmations below the threshold at rates of the same order as those above it.

Raising the NLI threshold from 0.5 to 0.9 removes at most 3 confirmed contradictions in each corpus (OPPT from 32 to 29, OPP-115 from 79 to 76), indicating that, among judged pairs, confirmed contradictions concentrate near the top of the NLI range rather than near the 0.5 flagging threshold. Lowering the threshold below 0.5 cannot change the result, since no pair below 0.5 was submitted for judging; this analysis is therefore one-sided and does not bound recall in the low-NLI region, where the segment-level pilot suggests most missed contradictions live.

\section{Cross-Temporal Comparison: OPP-115 Analysis}
\label{sec:cross_temporal}

To test whether contradiction patterns are structural or era-specific, we applied the pipeline to OPP-115~\citep{wilson2016creation} (115 companies; 3,792 segments, of which 2,878 were retained for analysis; policies collected in 2015). The primary finding is pattern consistency: the same dominant categories appear in both corpora. Prior longitudinal work documented policy evolution after the GDPR~\citep{linden2020privacy}; our comparison spans the pre-GDPR/CCPA environment (OPP-115) and the post-regulation landscape (OPPT).

\subsection{Scale Results}

Table~\ref{tab:comparison} compares results across the two corpora.

\begin{table}[htbp]
\centering
\begin{tabular}{lrr}
\toprule
\textbf{Metric} & \textbf{OPPT (60 judged)} & \textbf{OPP-115 (82 judged)} \\
\midrule
Statements extracted & 6,061 & 4,975 \\
Similarity-filtered pairs & 3,965 & 9,877 \\
Judged pairs & 293 & 663 \\
Panel-confirmed contradictions & 32 & 79 \\
Confirmation rate & 10.9\% & 11.9\% \\
Companies with contradictions & 15/60 (25\%) & 42/82 (51\%) \\
Fleiss' $\kappa$ & 0.48 & 0.57 \\
\bottomrule
\end{tabular}
\caption{Cross-temporal comparison of pipeline results. The two runs are not configuration-matched: OPPT applies metadata compatibility filters that the OPP-115 run does not (Section~\ref{sec:methodology}). The higher OPP-115 prevalence (51\% vs.\ 25\%) may therefore reflect substantive differences in privacy policy practices, corpus composition, or this filtering asymmetry. The judged-company rates are computed over companies with judged pairs and may overstate full-corpus prevalence (Section~\ref{sec:results}). Confirmation rates are comparable (11.9\% vs.\ 10.9\%). See text for analysis of alternative explanations.}
\label{tab:comparison}
\end{table}

The higher prevalence in OPP-115 has multiple candidate explanations, and the prevalence difference cannot be attributed to any single factor. The comparison's primary value lies in demonstrating pipeline applicability across both corpora and revealing consistent category patterns, not in measuring regulatory impact. Candidate explanations include:

\begin{itemize}
    \item \textbf{Corpus selection bias}: OPP-115 was designed for annotation research and may over-represent policies with diverse data practices.
    \item \textbf{Pre-regulation language}: Older policies may use less carefully lawyered language predating CCPA and GDPR compliance reviews.
    \item \textbf{Filtering asymmetry}: the OPP-115 run was configured without the subject, aspect, scope, and qualifier compatibility filters, including the LEGAL\_REQUIREMENT scope filter and qualifier-coverage filter that suppress legally mandated carve-outs and qualified commitments in OPPT. Regulatory-exception pairs such as Kaleida Health's HIPAA-permitted disclosures reach judges in OPP-115 but would likely have been filtered in OPPT. Because the OPP-115 statements carry the same metadata fields, a matched-configuration re-run is feasible; the stability experiment (Section~\ref{sec:stability}) performs one and finds that the matched gap narrows to roughly twofold, so this asymmetry explains part, but not all, of the difference observed here.
    \item \textbf{Category asymmetry}: SALE\_SHARING framing is more prevalent in OPPT, reflecting CCPA-era disclosure requirements, and contributes candidates with distinct linguistic characteristics.
\end{itemize}

\noindent Disentangling the substantive factors would require controlled experiments, though the filtering asymmetry in particular is removable with a matched-configuration re-run. The confirmation rates are nonetheless similar (11.9\% vs.\ 10.9\%) despite the differing filter configurations.

Statement typing is predominantly LLM-inferred in both corpora: the intended annotation guidance was inoperative for OPP-115 and minimal for OPPT (Section~\ref{sec:dataset}). The comparable confirmation rates therefore cannot be attributed to annotation provenance.

Processing both corpora required approximately 3 hours total at a combined API cost under \$20 across three providers. Metadata-based filtering reduces judge input to 956 pairs (293 + 663), enabling cost-effective analysis.

The companies with the most contradictions reflect domain-specific patterns that differ across regulatory eras. The leading OPPT companies (Duolingo, Microsoft, Google, and Roblox, 4 each) are discussed in Section~\ref{sec:results}. In OPP-115, Kaleida Health leads with 9 panel-confirmed contradictions, accounting for 11\% of all OPP-115 contradictions. We note this concentration as a potential confound in corpus-level analysis. Excluding Kaleida Health, the remaining OPP-115 prevalence remains substantial at 41 companies with contradictions out of 81 judged (50.6\%), and the total contradiction count remains high at 70. This suggests the overall pattern is not driven by this single outlier. These contradictions reflect systematic tension between HIPAA authorization requirements and legally-permitted exceptions:

\begin{quote}
\small
\textbf{COMMITMENT:} Kaleida Health obtains written authorization before using or sharing health information with external parties.

\textbf{PRACTICE:} Kaleida Health uses and discloses patient health information in the Patient Directory without written authorization.

\textbf{Verdict:} UNANIMOUS (similarity=0.86, NLI=1.00)
\end{quote}

These contradictions reflect HIPAA compliance requirements (45 C.F.R. \S~164.520 mandates enumeration of permitted exceptions~\citep{hipaa164520}) yet, as discussed in Section~\ref{sec:limitations}, remain textual contradictions. HIPAA-mandated contradictions may be categorically different from CCPA-related ones, arising from a regulatory structure requiring simultaneous statement of rules and exceptions rather than accretion of new disclosures, and warrant further investigation with a dedicated healthcare corpus.

IMDB, Allstate, and Zacks each have 4 contradictions. ABCNews, USA.gov, JibJab, Time Inc., and Reddit each have 3. The remaining 33 companies with contradictions each have 1--2 confirmed cases. Per-company OPP-115 results parallel to Table~\ref{tab:oppt_companies} are available in the released artifacts (see Data and Code Availability).

\subsection{Pattern Consistency}

The same dominant category patterns appear in both corpora, consistent with structural rather than company-specific causes. Table~\ref{tab:opp115_categories} presents the OPP-115 category distribution for comparison with the OPPT distribution in Table~\ref{tab:categories}. THIRD\_PARTY-to-THIRD\_PARTY contradictions dominate OPP-115 (40.5\% of panel-confirmed contradictions) and remain the joint-largest pattern in OPPT (18.8\%). FIRST\_PARTY-to-FIRST\_PARTY contradictions appear in both eras (24.1\% in OPP-115, 18.8\% in OPPT).

\begin{table}[htbp]
\centering
\begin{tabular}{lrr}
\toprule
\textbf{Commitment Category $\to$ Practice Category} & \textbf{Count} & \textbf{\%} \\
\midrule
THIRD\_PARTY $\to$ THIRD\_PARTY & 32 & 40.5\% \\
FIRST\_PARTY $\to$ FIRST\_PARTY & 19 & 24.1\% \\
FIRST\_PARTY $\to$ THIRD\_PARTY & 10 & 12.7\% \\
SALE\_SHARING $\to$ THIRD\_PARTY & 9 & 11.4\% \\
THIRD\_PARTY $\to$ FIRST\_PARTY & 4 & 5.1\% \\
Other & 5 & 6.3\% \\
\bottomrule
\end{tabular}
\caption{Category pair distribution of panel-confirmed OPP-115 contradictions. Third-party sharing contradictions dominate at 40.5\%, consistent with the OPPT pattern (Table~\ref{tab:categories}).}
\label{tab:opp115_categories}
\end{table}

The absolute similarity distributions of panel-confirmed contradictions look similar across corpora (32.9\% of OPP-115 and 40.6\% of OPPT confirmations fall in the 0.50--0.55 range), but the confirmation rates by bin diverge: OPP-115 rates rise with similarity (9.7\% at 0.50--0.55, 19.3\% at 0.65--0.70, 27.3\% above 0.75), while OPPT rates are flat with sparse high-similarity cells (Section~\ref{sec:results}). We therefore do not interpret the similarity distribution as evidence about the nature of privacy washing.

The SALE\_SHARING-to-THIRD\_PARTY pattern appears more often in OPPT (18.8\%, 6 of 32) than in OPP-115 (11.4\%, 9 of 79), though the difference is not statistically significant (two-sided Fisher's exact test, $p = 0.36$). The direction is consistent with CCPA-mandated ``we do not sell'' disclosures added after January 2020, though sale-related commitments also appear in the pre-CCPA corpus, and the mechanism argument for regulatory accretion (Section~\ref{sec:discussion}) does not rest on this comparison. These statements, prompted in the post-CCPA era by Cal. Civ. Code \S~1798.130(a)(5)(C)'s requirement that businesses either list the categories of personal information sold or shared or prominently disclose that no sale or sharing occurred~\citep{ccpa1798130}, create contradictions when companies maintain pre-existing sharing practices that may not meet the legal definition of ``sale'' but nonetheless involve extensive data disclosure.

The 11-year gap between corpora spans a transformative regulatory period. OPP-115 policies predate the GDPR (effective May 2018) and the CCPA (effective January 2020). The persistence of THIRD\_PARTY contradiction patterns in both eras suggests structural features of policy composition. The greater prevalence of SALE\_SHARING contradictions in OPPT is consistent with CCPA's mandated ``do not sell'' disclosures creating new contradiction opportunities.

\subsection{Prevalence Across Corpora}

The primary finding from the cross-corpus comparison is pattern consistency: the same dominant contradiction categories appear in both corpora (Section~\ref{sec:cross_temporal}), suggesting structural rather than era-specific causes. Across both corpora, 57 of 142 judged companies (40\%) have at least one panel-confirmed contradiction among those reaching the judge panel. Measured rates differ between corpora: 25\% in OPPT (15/60 judged; 95\% Clopper--Pearson CI: 15--38\%) and 51\% in OPP-115 (42/82 judged; 95\% Clopper--Pearson CI: 40--62\%). A two-sided Fisher's exact test finds this difference statistically significant ($p = 0.002$, odds ratio $= 0.32$), but the test compares two differently configured pipeline runs (Section~\ref{sec:methodology}) as much as two corpora, so the result should not be read as measuring a corpus or era difference. Full-corpus prevalence is 12.2\% (15/123; 95\% Clopper--Pearson CI: 7.0--19.3\%) for OPPT as run (9.8\%, 12/123, in the reproducible subset) and 36.5\% (42/115; 95\% CI: 27.7--46.0\%) for OPP-115. These figures are lower bounds on true prevalence: the 63 unjudged OPPT companies (45 with no NLI-flagged pairs, 18 with flagged pairs below the judge-submission similarity threshold) and 33 unjudged OPP-115 companies (19 and 14 respectively) were never evaluated by judges and may contain contradictions the pipeline's filtering stages did not surface.

Absence of a panel-confirmed contradiction may reflect a simpler policy, sector-specific drafting conventions, or the pipeline's filtering; we do not characterize any company's absence of findings as evidence of stronger privacy practices. One category, companies that avoid contradictions through commitment avoidance rather than practice limitation, is analyzed in Section~\ref{sec:commitment_avoidance}.

\subsection{Commitment Avoidance}
\label{sec:commitment_avoidance}

A distinct pattern emerges among companies with zero panel-confirmed contradictions. Practice-to-commitment ratios depend on the accuracy of upstream COMMITMENT/PRACTICE typing, which has not been validated against human labels; these observations are preliminary indicators.

Amazon, Netflix, and Airbnb each had zero panel-confirmed contradictions; all 11 judged pairs from these companies were unanimously rejected by judges. These companies exhibit high practice-to-commitment ratios in our extraction output: Netflix at 4.7:1 and Airbnb at 10.3:1, with only 3 total commitment statements extracted from Airbnb's entire policy, a count low enough that manual verification of the extraction would be warranted before interpreting the ratio. Amazon's scope qualifiers are narrowly constructed (e.g., ``when providing to ad companies,'' an author-selected illustration from the extraction output, not a panel-confirmed pair), avoiding logical contradiction while describing extensive data practices. A policy with few affirmative privacy promises has structurally less exposure to contradiction detection than one with many; whether these ratios reflect drafting strategy, business model, or extraction behavior cannot be determined from these data.

High practice-to-commitment ratios also appear among companies with panel-confirmed contradictions: Uber (3 contradictions) exhibits an 18.2:1 ratio and Meta (1 contradiction) 9.0:1, while among companies with zero panel-confirmed contradictions, Lyft exhibits 7.0:1 and T-Mobile 6.7:1, though with only 3 and 17 judged pairs respectively, these two absences establish little. The highlighted companies are extreme outliers, not typical: across all 123 OPPT companies (every one of which has at least one commitment statement), the median ratio is 1.71 (interquartile range 1.24--2.51), only 5 companies exceed 5:1, and only 2 exceed 10:1. At the corpus level, the ratio is not significantly associated with having a panel-confirmed contradiction (Spearman $\rho = 0.146$, $p = 0.107$, $n = 123$, against a binary indicator of post-reclassification panel-confirmed contradiction). This test is confounded by policy length (longer policies produce more judged pairs and hence more confirmation opportunities) and has power to detect only large effects with 15 positive companies among 123. The hypothesis that commitment avoidance substitutes for commitment violation is therefore not detectable at this sample size; whether the ratio could serve as a complementary metric, measuring commitment avoidance alongside commitment violation, remains future work.

\section{Stability Experiment: Separated Panels and Matched Configuration}
\label{sec:stability}

Seven months after the primary runs, we re-ran the complete pipeline on both corpora (August 30--31, 2026) with a configuration designed to test three of the paper's acknowledged weaknesses at once. The primary results reported above are unchanged by this experiment; this section reports how they behave under a substantially different instantiation of the same design.

\subsection{Design}

The stability run differs from the primary runs in four ways. First, the extraction and judge panels are fully separated: extraction uses three models from Western providers (Claude Haiku 4.5, GPT-5.6 Luna, Gemini 3.7 Flash, the successors to the primary panel's models), while judging uses three models from Chinese providers (DeepSeek V4 Flash, GLM-5.3-Flash, Kimi K3). No model extracts and judges the same statements, removing the extractor-judge overlap concern (Section~\ref{sec:limitations}), and none of the judge providers appears as a company in either corpus, removing the judge conflict of interest (Section~\ref{sec:ethics}). Second, the metadata compatibility filters are enabled for both corpora, placing the cross-corpus comparison on the matched configuration the primary runs lacked. Third, the judge stage evaluates every NLI-flagged pair rather than only those at or above the 0.5 similarity threshold, closing the recall region the sensitivity analysis (Section~\ref{sec:sensitivity}) could not reach. Fourth, fresh experiment directories guarantee an empty judge cache, so no legacy pairs can enter. The judge stage cost \$12.30 in API fees across the three providers (2,011 pairs; extraction cost was not separately metered).

\subsection{Results}

Table~\ref{tab:stability} compares the stability run against the primary runs. Because extraction changed along with the judges, differences conflate extractor and judge effects; the comparison tests the pipeline design as a whole, not the judge stage in isolation.

\begin{table}[htbp]
\centering
\begin{tabular}{lrrrr}
\toprule
& \multicolumn{2}{c}{\textbf{OPPT}} & \multicolumn{2}{c}{\textbf{OPP-115}} \\
\textbf{Metric} & \textbf{Primary} & \textbf{Stability} & \textbf{Primary} & \textbf{Stability} \\
\midrule
Statements extracted & 6,061 & 7,082 & 4,975 & 6,485 \\
NLI-flagged pairs & 704 & 834 & 1,781 & 1,177 \\
Judged pairs & 293 & 834 & 663 & 1,177 \\
\quad at similarity $\geq$ 0.5 & 293 & 348 & 663 & 422 \\
Confirmed (similarity $\geq$ 0.5) & 32 & 29 & 79 & 58 \\
Confirmation rate ($\geq$ 0.5) & 10.9\% & 8.3\% & 11.9\% & 13.7\% \\
Confirmed (all judged pairs) & --- & 56 & --- & 155 \\
Full-corpus prevalence ($\geq$ 0.5) & 12.2\% & 13.0\% & 36.5\% & 27.8\% \\
Full-corpus prevalence (all) & --- & 20.3\% & --- & 42.6\% \\
Unanimous verdicts & 83.6\% & 91.4\% & 81.4\% & 86.4\% \\
Fleiss' $\kappa$ & 0.48 & 0.57 & 0.57 & 0.64 \\
\bottomrule
\end{tabular}
\caption{Primary runs (January--February 2026, as-run figures, post-reclassification) versus the stability run (August 2026, pre-audit figures; the parallel reclassification audit is reported in the text and changes four OPP-115 stability figures: 155 to 140 confirmed, 58 to 51 at $\geq$ 0.5, 27.8\% to 25.2\%, and 42.6\% to 40.9\%; the OPPT changes are 56 to 54 and 29 to 27, with no prevalence change). The stability run's OPP-115 pair counts reflect the matched metadata filtering absent from the primary OPP-115 run, and its judged set includes all NLI-flagged pairs, with the similarity $\geq$ 0.5 subset recovering the primary protocol.}
\label{tab:stability}
\end{table}

Five findings emerge. First, prevalence is stable under the primary protocol. At the matched judge-submission threshold, OPPT full-corpus prevalence is 13.0\% (16/123) against the primary 12.2\% (15/123), and the confirmation rate is 8.3\% against 10.9\%, under a different extraction panel, a judge panel from three different providers, and an empty cache. This matched-protocol claim is OPPT-only: the OPP-115 comparison (27.8\% vs.\ 36.5\%) additionally changes the filter configuration, so its difference conflates panel and filter effects. Judge agreement is higher in the stability run for both corpora (Fleiss' $\kappa$ 0.57 vs.\ 0.48 for OPPT, 0.64 vs.\ 0.57 for OPP-115), and the per-judge conservatism spread recurs with different models in the roles: GLM-5.3-Flash votes CONTRADICTION on 4.6\% of OPPT pairs against 10.3\% for DeepSeek, mirroring the primary panel's Google-versus-OpenAI spread.

Second, the legacy-pair findings fail as the primary analysis predicted. Google, Appriss, Bumble, and Tesla, the four companies whose primary counts depended on legacy pairs (Table~\ref{tab:oppt_companies}), produce no confirmed contradiction in the stability run, an outcome partly guaranteed by construction: the run both starts from an empty judge cache and retains the qualifier-coverage filter that excludes pairs of the legacy class, so the residual information is only that the new extraction produced no differently phrased statements that pass the filter for these companies. Of the 12 companies in the primary reproducible subset, 8 recur (Duolingo, GitHub, Meta, Microsoft, Roblox, Uber, Venmo, Walmart), as does Khan Academy from the legacy-dependent set; Notion, Motorola Solutions, and the remaining legacy-dependent companies do not. For OPP-115, 34 of the 46 pre-reclassification primary companies recur somewhere in the stability run's confirmations (27 at similarity $\geq$ 0.5, 7 below it); of the 12 losses, 9 had pairs judged and rejected and 3 lost their pairs to the matched filters upstream of judging.

Third, the sub-threshold region contains confirmations at rates of the same order as the region at or above the threshold, converting the paper's lower-bound language into a measured quantity. Below the 0.5 similarity threshold, OPP-115 pairs confirm at 12.8\% (97/755) against 13.7\% above it (two-sided Fisher's exact test, $p = 0.65$); OPPT pairs confirm at 5.6\% (27/486) against 8.3\% above, a directional but not statistically significant gap ($p = 0.12$). Removing the threshold raises full-corpus prevalence to 20.3\% for OPPT and 42.6\% for OPP-115 (40.9\% after the reclassification audit below). In OPP-115, 63\% of all confirmations lie below the threshold the primary runs never judged.

Fourth, the cross-corpus gap survives the matched configuration. With identical filters, OPP-115 full-corpus prevalence (27.8\% at the matched threshold) remains roughly twice OPPT's (13.0\%), against the roughly threefold gap in the unmatched primary runs. The filtering asymmetry therefore explains part, but not all, of the primary gap; the residual is consistent with a real corpus or era difference, though corpus selection and pre-regulation drafting (Section~\ref{sec:cross_temporal}) remain unseparated candidate explanations.

Fifth, the category patterns recur, but their ordering does not. The same small set of category pairs dominates both corpora in both runs. Under the stability panel, however, FIRST\_PARTY $\to$ FIRST\_PARTY is the modal pair in both corpora (30 of 56 OPPT confirmations, 84 of 155 OPP-115), and pairs with third-party sharing on either side fall to 43\% in each corpus, below the majority reported for the primary runs (56.3\% and 69.6\%). The primary abstract-level claim that third-party sharing accounts for the majority of confirmed cases is thus panel-sensitive, while the weaker claim, that the same handful of first-party and third-party category pairs recurs across corpora and eras, is what survives reconfiguration. The shift also bears on the prompt-example priming concern (Section~\ref{sec:limitations}): the judge prompt's examples all involve third-party or sale patterns, and this panel, given the same prompt, confirmed proportionally fewer of them.

\subsection{Reclassification Audit and Caveats}

We applied the same three-class reclassification audit used on the primary runs (Section~\ref{sec:limitations}) to the stability run's confirmations, classifying each unique commitment-side statement with the stability run's extraction panel. For OPPT, 2 of 56 confirmations (3.6\%) involve a reclassified commitment (both retyped as PRACTICE), closely matching the primary run's 3.0\%; the corrected total is 54, and no company count changes. For OPP-115, 15 of 155 confirmations (9.7\%) involve reclassified commitments (10 statements retyped as PRACTICE, 2 as USER\_CONTROL), roughly half the primary run's 20.2\% affected rate, consistent with the attribution of that asymmetry to the primary OPP-115 run's missing metadata filters. The corrected OPP-115 totals are 140 confirmations, 47 companies (40.9\%) over all judged pairs, and 51 confirmations across 29 companies (25.2\%) at the matched threshold. (The stability run also had 3 OPP-115 pairs, 0.3\%, without a usable verdict, from split two-judge outcomes after one invalid response; OPPT had none.) No corrected figure changes any conclusion above: the OPPT matched-threshold prevalence remains 13.0\% (16/123), the sub-threshold and above-threshold confirmation rates remain of the same order within each corpus (OPPT 5.6\% vs.\ 7.8\%; OPP-115 11.8\% vs.\ 12.1\% after correction), and the matched-configuration cross-corpus gap remains roughly twofold. The category conclusion of finding 5 also survives the correction under a worst-case bound: even if all 15 removed OPP-115 confirmations were FIRST\_PARTY $\to$ FIRST\_PARTY, that pair would remain modal (69/140 vs.\ 52/140 for THIRD\_PARTY $\to$ THIRD\_PARTY) and the third-party share would remain below a majority (at most 67/140, 47.9\%).

The stability run does not touch the paper's central limitation: none of its verdicts have been validated against human judgment, and agreement between two LLM panels is not accuracy. The run also changes extraction and judging together, so company-level churn (16 OPPT companies confirm only in the stability run, largely from the enlarged judged set and the larger extraction yield) cannot be attributed to either stage alone. All stability-run artifacts, including the failed first judge pass and per-judge verdicts, are released alongside the primary artifacts (see Data and Code Availability).

\section{Discussion}
\label{sec:discussion}

\subsection{Structural Factors Contributing to Privacy Washing}
\label{sec:structural}

The contradictions we observe are consistent with identifiable structural factors in how privacy policies are written and maintained, not necessarily with intentional deception. We identify five such factors: regulatory accretion, multi-division policies, acquisition boilerplate, template evolution, and commitment asymmetry. We present these as hypothesized mechanisms grounded in qualitative examination of confirmed cases, not as quantified findings: we did not code each confirmed pair by mechanism, and one factor (acquisition boilerplate) is not represented among the confirmed contradictions of the enhanced pipeline.

Regulatory accretion contributes to contradiction patterns. The California Consumer Privacy Act requires companies to disclose whether they sell or share personal information and to honor opt-out rights~\citep{ccpa1798120,ccpa1798130,ccpa1798135}, and companies add ``we do not sell'' claims to comply. These broad statements conflict with pre-existing descriptions of sharing practices that may not meet the legal definition of ``sale'' but nonetheless involve extensive data disclosure.

The prevalence of SALE\_SHARING contradictions reflects a structural gap in privacy law. CCPA defines ``sale'' as disclosure ``for monetary or other valuable consideration''~\citep[\S~1798.140(ad)]{ccpa1798140} but leaves ``valuable consideration'' undefined. In the 2022 Sephora settlement~\citep{sephora2022settlement}, the California Attorney General took the position that receiving advertising services in exchange for consumer data qualifies, yet companies continue claiming ``we do not sell'' while engaging in such exchanges. The California Privacy Rights Act~\citep{cpra2020} partially addressed this by adding an explicit ``sharing'' concept for cross-context behavioral advertising, but the original CCPA's definitional ambiguity created the structural conditions for the SALE\_SHARING pattern we observe.

Multi-division policies amplify contradictions by placing product-specific commitments alongside enterprise-wide practices. Microsoft's policy covers Windows, Xbox, Office, Bing, LinkedIn, and dozens of other services, where a commitment that ``Phone Link does not store app data'' (an author-selected illustration from the policy text, not a panel-confirmed pair) appears accurate for that specific product yet sits alongside enterprise-wide collection of app names, versions, and activity data. This pattern suggests that organizational structure contributes to contradictions.

Acquisition boilerplate may introduce systematic contradictions. Standard clauses granting acquiring entities the right to continue using customer data create tension with privacy commitments such as no-sharing-without-permission and user notification of material changes. In preliminary experiments with less restrictive filtering, acquisition clauses generated numerous NLI-flagged pairs, though most were rejected by judges under the enhanced pipeline's stricter criteria; this factor therefore remains a hypothesis rather than a finding.

Template evolution contributes to contradictions when policies grow by accretion over years, with new sections added without reconciling against existing commitments. Unlike regulatory accretion, which introduces contradictions through specific mandated disclosures, template evolution operates through ordinary maintenance: successive edits by different drafters accumulate practice descriptions that no one checks against commitments made in earlier revisions, so the contradiction is a by-product of document history rather than of any single drafting decision.

Commitment asymmetry functions as a cross-cutting factor: user-facing commitments employ broad, reassuring language (``we do not share your data'') while practice descriptions use legally precise language (``we disclose identifiers, commercial information, and internet activity to service providers, affiliates, and partners''). Both may be accurate on their own terms, but the commitment creates an impression that the practice undermines.

\subsection{Why Structured Metadata Matters}
\label{sec:metadata}

The three false positive patterns identified in preliminary experiments (subject mismatches, aspect mismatches, qualifier losses) motivated the structured metadata schema, whose compatibility filters remove 71.9\% of category-compatible OPPT pairs before similarity and NLI scoring. The demonstrated value of these filters is cost: they eliminate most candidate pairs before expensive judge verification. The paper contains no filter ablation. The OPP-115 run was configured without the metadata filters, but it also differs in corpus, era, and candidate distribution, so the difference in confirmation rate (11.9\% versus 10.9\%) cannot be attributed to the filters; a matched-configuration ablation, computable from the released artifacts without re-running extraction, remains future work (Section~\ref{sec:future}). The one piece of within-corpus evidence is unfavorable to the qualifier-coverage filter: the 30 legacy pairs, which fail exactly that filter, were confirmed by judges at 26.7\% versus 9.5\% for retained pairs (Section~\ref{sec:methodology}), so the filter discards a class of pairs the judges disproportionately confirm. Field-level pairwise agreement rates across the three-LLM consensus panel vary by dimension: subject 82\%, type 74\%, aspect 69\%, scope 65\%. These are raw pairwise rates over fields with skewed marginals (most subjects are COMPANY, most scopes UNIVERSAL), so chance-corrected agreement would be lower. The aspect and scope fields, with the lowest agreement, drive the two largest metadata filters (12,758 and 578 pairs removed respectively), meaning a substantial share of filtering decisions rest on labels the extractors themselves only moderately agree on; a controlled filter ablation on matched configurations is future work.

\subsection{The Complementarity of NLI and Judge Signals}

The NLI signal errs in both directions. The judge rejection rates among judged pairs (89.1\% for OPPT, 88.1\% for OPP-115) align with concerns about spurious correlations in NLI~\citep{mccoy2019right,gururangan2018annotation} and performance degradation on legal text; the judge panel captures pragmatic reasoning NLI cannot, including hedging language, conditional scope, product boundaries, and data type distinctions. In the other direction, the segment-level pilot found that 89\% of judge-confirmed pairs were never NLI-flagged (Section~\ref{sec:pilot}), so NLI screening also bounds the pipeline's recall from above. NLI is thus useful as an inexpensive screen but is neither a reliable precision signal nor a reliable recall signal on this task without downstream verification. Confirmation rates should be interpreted with the extractor-judge overlap (Section~\ref{sec:methodology}) in mind.

\subsection{Comparison with PolicyLint}

PolicyLint~\citep{andow2019policylint} and our approach detect complementary classes of contradiction. PolicyLint identifies logical negation within structured data-collection tuples using symbolic NLP and ontologies, requiring no language models but limited to contradictions that can be expressed as tuple inversions. Our approach detects pragmatic contradictions between commitments and practices using metadata-enhanced extraction, NLI screening, and multi-model judge verification.

PolicyLint would not flag ``we do not sell'' paired with ``we share identifiers with advertising partners'': ``sell'' and ``share'' are different actions with no tuple negation. Our approach identifies the pragmatic contradiction: the broad reassurance creates an impression that the practice undermines.

The prevalence findings differ accordingly. PolicyLint found contradictions in 14.2\% of 11,430 app policies, while our full-corpus rates are 12.2\% of OPPT companies and 36.5\% of OPP-115 companies. These figures are not directly comparable: they differ in contradiction definition (pragmatic vs.\ logical), unit of analysis (companies vs.\ individual policies), corpus scale (238 vs.\ 11,430), and policy type (website vs.\ mobile app).

The two approaches are complementary rather than competing. A comprehensive policy audit would benefit from both: PolicyLint-style analysis for logical consistency of specific data claims, and our approach for detecting the pattern of reassuring commitments undermined by documented practices. The enhanced metadata extraction (subject, aspect, scope, qualifiers) could be applied to improve PolicyLint-style systems as well. These complementarity claims are analytical, grounded in the systems' designs; we did not run PolicyLint or PoliGraph on our corpora, and measuring the empirical overlap remains future work (Section~\ref{sec:limitations}).

\subsection{Commitment Avoidance, Vagueness, and Detection Limits}
\label{sec:vagueness}

An important caveat concerns the nature of ambiguity itself. We distinguish between \emph{strategic vagueness} (deliberate use of imprecise terms to provide flexibility) and \emph{drafting errors} that create unintended contradictions. \citet{reidenberg2016ambiguity} found that companies use ambiguity when summarizing their data practices to preserve flexibility for future practices, identifying four categories of vagueness: conditionality, generalization, modality, and numeric quantifiers. \citet{bhatia2016vagueness} developed a taxonomy of vague terms (``may,'' ``as necessary,'' ``generally'') that companies deploy systematically, and \citet{lebanoff2018automatic} demonstrated automatic detection of vague words and sentences in privacy policies, an instrument directly applicable to the commitment-avoidance measurement discussed below.

Our detection pipeline cannot distinguish between intentional strategic vagueness and unintentional contradiction. A commitment stating ``we may share data with partners'' followed by ``we share data with advertising partners'' is not a contradiction; the hedging language explicitly permits the practice. However, a commitment stating ``we do not sell your personal information'' followed by ``we disclose identifiers to advertising partners'' may represent either (a) intentional gap-exploitation using narrow legal definitions, (b) unintentional contradiction from template evolution, or (c) deliberate privacy washing. The pipeline detects the textual contradiction but cannot determine which category applies.

Whether vagueness and commitment avoidance represent distinct policy failure modes complementary to privacy washing, or merely detection limits, warrants separate investigation. The practice-to-commitment ratio observations in Section~\ref{sec:commitment_avoidance} (Netflix 4.7:1, Airbnb 10.3:1, Amazon with narrow scope qualifiers) illustrate the hypothesis that commitment avoidance may be as policy-relevant as commitment violation, but the corpus-level association between the ratio and confirmed contradictions is null, and measuring commitment avoidance would require different methods than contradiction detection.

This limitation is inherent to automated detection. Future work should explore whether contradiction patterns correlate with proxies for intentionality: temporal ordering of section additions, organizational complexity, or industry-specific regulatory pressures. Human expert review with access to policy revision history could provide ground truth on the intentional/unintentional distinction.

\subsection{Implications}

Before discussing applications, we note that the author has a commercial interest in privacy policy analysis technology (see the Competing Interests statement); the claims in this subsection should be read with that interest in mind. Following expert validation of precision, privacy washing detection could provide regulators with a screening indicator aligned with existing enforcement standards. Under the FTC's ``net impression'' doctrine~\citep{ftc1983deception}, even literally true statements can be deceptive when their overall impression misleads, and internal contradiction is a textual condition under which such impressions arise. A pipeline that detects internal contradictions could identify candidates for expert review, enable continuous monitoring of published policies, and provide a basis for sector-wide compliance assessments. The finding that CCPA ``no sell'' disclosures are among the most commonly contradicted commitment types is directly relevant to the California Privacy Protection Agency's (CPPA) enforcement priorities.

Recent CPPA enforcement shows regulators acting against gaps between ``do not sell'' representations and actual data flows: the September 2025 Tractor Supply settlement~\citep{cppa2025tractorsupply} (\$1.35 million) targeted a ``Do Not Sell My Personal Information'' link that appeared functional but failed to stop data sales via website trackers. That case concerned a behavioral gap between an interface control and backend behavior, which our text-only pipeline could not have detected; internal textual contradiction is a related but distinct signal that could surface candidates for such investigations. The CPPA's September 2024 dark patterns advisory explicitly adopts an ``effect not intent'' standard~\citep{cppa2024darkpatterns}, aligning with our structural approach to privacy washing.

The FTC's enforcement actions have created what \citet{solove2014ftc} characterize as a ``new common law of privacy,'' establishing principles through consent decrees and settlements rather than statutory definitions. The deception framework remains active for privacy representations: the FTC's 2023 GoodRx action turned on gaps between a company's privacy representations and its actual data flows~\citep{ftc2023goodrx}, and the agency's 2022 staff report on dark patterns documents design practices that obscure or contradict material privacy information~\citep{ftc2022darkpatterns}. Under the ``net impression'' doctrine (Section~\ref{sec:definition}), even literally true claims may be deceptive if they create a misleading impression. Privacy washing creates precisely this condition: a commitment such as ``we do not sell your personal information'' may be literally true under CCPA's narrow definition of ``sale,'' yet the documented practice of disclosing identifiers and commercial information to advertising partners creates a net impression that contradicts reasonable consumer expectations.

Pipeline outputs should be understood as candidates for expert review, not as automated determinations of legal violation, and we do not recommend deployment of this pipeline for enforcement purposes prior to expert validation of precision, a caution consistent with the broader literature on the risks of deploying privacy-policy NLP for consumer-facing and enforcement purposes~\citep{ravichander2021breaking}. We distinguish three categories of findings. \emph{Strong prima facie candidates} include SALE\_SHARING commitments contradicted by third-party practices (18.8\% of OPPT contradictions), which resemble the fact pattern in the Sephora settlement, where receiving advertising services constituted ``valuable consideration.'' \emph{Patterns requiring regulatory guidance} include acquisition clause contradictions and product-boundary contradictions in multi-division policies, where existing precedent does not clearly resolve whether the contradiction is actionable. \emph{Structural indicators for policy improvement} include commitment-asymmetry cases, where a broad reassuring commitment coexists with a legally precise practice description (Section~\ref{sec:structural}); these may not constitute violations but indicate conditions for consumer confusion warranting voluntary remediation.

For companies, the pipeline could support self-audit following validation. Running the detection pipeline before publishing policy updates would surface candidate inconsistencies before they reach users. The metadata-based filtering reduces the volume of candidate pairs requiring review, though its effect on precision is untested (Section~\ref{sec:metadata}). The structural patterns we identify (regulatory accretion, multi-division fragmentation, acquisition boilerplate) suggest specific remediation strategies: reconcile new regulatory disclosures against existing practice descriptions, audit product-specific commitments against enterprise-wide collection, and review standard M\&A clauses for conflicts with stated privacy guarantees.

For privacy researchers, privacy washing represents a dimension of policy quality distinct from readability. Prior work establishes that most users skip policies~\citep{obar2020biggest}, navigate by section headings~\citep{vu2007users}, hold expectations misaligned with actual policy contents~\citep{reidenberg2015disagreeable}, essentially never read standard-form contracts~\citep{bakos2014fineprint}, and are only weakly influenced by variations in policy language~\citep{strahilevitz2016irrelevant}, the last a finding in tension with the premise that policy language shapes consumer expectations. These findings are consistent with, but do not establish, a scenario in which a reader encounters the commitment without reaching the contradicting practice.

\medskip
\noindent Combined with evidence of jurisdiction-siloed disclosure~\citep{brackin2026jurisdiction}, effectiveness limitations of short-form notices~\citep{gluck2016short}, and increasing use of pacifying language~\citep{belcheva2023understanding}, these results suggest that policy architecture and consistency deserve attention alongside policy language and readability.

\section{Limitations}
\label{sec:limitations}

Several limitations warrant discussion. The overarching one is that no stage of the pipeline (extraction, statement typing, or judging) has been validated against human labels; every reported quantity is a property of the pipeline, not a measurement of contradiction prevalence. The stage-specific concerns below should be read against that cumulative gap. Our findings span 238 companies across two corpora, and larger studies across additional jurisdictions would further test generalizability.

We did not directly validate COMMITMENT/PRACTICE typing, which is predominantly LLM-inferred in both corpora (Section~\ref{sec:dataset}); the 15.9\% reclassification rate among confirmed pairs shows it is imperfect. Inter-annotator agreement studies on extracted statements remain important future work.

A refined three-class taxonomy (COMPANY\_COMMITMENT, PRACTICE, USER\_CONTROL) revealed that the binary COMMITMENT label conflates company commitments with user capability statements~\citep{mysoresathyendra2017identifying}. For example, ``users can opt out of targeted advertising'' describes a user capability, not a company commitment. The audit described in Section~\ref{sec:methodology} removed the 21 affected confirmations (15.9\% of the original 132), and the removals fall disproportionately on the OPP-115 run, which lacked metadata filtering.

A primary methodological concern is that the same three models (Claude Haiku 4.5, GPT-5 mini, Gemini 3 Flash Preview) serve as both extractors (Stage 1) and judges (Stage 3). If a model systematically sharpens hedged language during extraction (e.g., converting ``we may share'' to ``we share''), it may be predisposed to confirm the resulting pair as contradictory during judging. LLM evaluators are known to recognize and favor their own generations~\citep{panickssery2024llm} and to exhibit systematic evaluation biases~\citep{wang2024unfair,zheng2023judging}. The tasks are functionally distinct (structured schema extraction vs.\ open-ended pragmatic evaluation), but shared model biases could propagate through both stages. The stability run (Section~\ref{sec:stability}) addresses this concern at the aggregate level: with fully separated extraction and judge panels, prevalence under the matched protocol is essentially unchanged, though the category composition of confirmations shifts, so the invariance does not extend to individual verdicts.

Results may shift as models improve. Even at temperature 0.0, LLM outputs can vary due to model updates and non-deterministic decoding; we ran the primary experiments once per corpus (OPPT in January 2026, OPP-115 in February 2026) and release all intermediate outputs for reproducibility. The stability run (Section~\ref{sec:stability}) is a second full run of each corpus, but under a deliberately different configuration, so run-to-run variation under an identical configuration remains unquantified; the repeated-judging experiment proposed in Section~\ref{sec:future} would measure it. All reported intervals are binomial and do not account for that variation. ``Panel-confirmed'' denotes LLM majority agreement, not expert validation; human annotation remains necessary future work.

The judge prompt's contradiction examples (Appendix~\ref{appendix:judge}) all involve third-party sharing or sale patterns, which could prime judges toward these categories. SALE\_SHARING commitments appear in 9 of 32 confirmed OPPT contradictions (28\%) and 10 of 79 OPP-115 contradictions (13\%). The base-rate analysis in Section~\ref{sec:results} shows that the confirmation-rate lift for third-party sharing pairs over other judged pairs is small and not statistically significant, so most of the category concentration arises upstream of the judges; that analysis cannot, however, rule out prompt-example effects at the judge stage. Because SALE\_SHARING framing is itself more prevalent in the post-CCPA OPPT corpus, however, the comparison cannot fully separate genuine CCPA-driven patterns from prompt-example bias; testing prompts with examples spanning multiple categories remains necessary future work. The prompt's instruction that statements ``about different products or services'' are not contradictions may also suppress detection of multi-division contradictions identified as genuine structural patterns in Section~\ref{sec:discussion}, and confirmed pairs from Duolingo and Uber (product- or group-specific commitments paired with general practice statements; Section~\ref{sec:results}) show that judges did not always apply this instruction conservatively.

Author review found no obvious false positives, but single-expert review does not constitute rigorous validation; multiple experts annotating blind to LLM verdicts~\citep{wilson2016creation} would be necessary to compute precision and recall.

Thresholds were selected from preliminary observations rather than systematic optimization. Section~\ref{sec:sensitivity} shows stable confirmation rates across variations, but judge verdicts exist only for pairs passing original thresholds, so recall below those thresholds is unknown.

The pipeline prioritizes precision over recall. The recall cascade begins at extraction: the Stage 1 consensus requirement discards statements produced by only one model, and extraction recall against source text is unmeasured, making this the only filter in the cascade whose loss rate is entirely unknown. Downstream, multiple filtering stages each discard candidates, some of which may be true contradictions, particularly those expressed through substantially different vocabulary, requiring multi-hop inference, or embedded in tables rather than prose. The 63 unjudged OPPT companies and 33 unjudged OPP-115 companies represent a systematic blind spot, as do the NLI-flagged pairs below the judge-submission similarity threshold (441 in the final OPPT pair set, 1,118 in OPP-115; the stability run judges the corresponding region of its own pair set and confirms pairs there at rates of the same order, Section~\ref{sec:stability}) and the segment-level pilot evidence that judges dismiss pairs a human reviewer classified as genuine (Section~\ref{sec:pilot}). The pipeline's recall is also bounded above by NLI recall, which the pilot suggests is low: 89\% of judge-confirmed pilot pairs were unflagged by NLI. Without a gold-standard corpus, quantifying recall remains infeasible.

No existing contradiction-detection system was run on these corpora as a baseline. The complementarity argument with PolicyLint (Section~\ref{sec:discussion}) is therefore analytical rather than empirical; running PolicyLint or PoliGraph on the OPPT corpus and measuring the overlap between their findings and ours is planned future work.

The SENSITIVE\_DATA category, which we introduced to reflect post-2016 regulatory developments, produced no panel-confirmed contradictions: its 67 OPPT statements yielded 28 judged pairs with zero confirmations, and the OPP-115 extraction emitted no SENSITIVE\_DATA statements at all. We cannot distinguish among three explanations (genuinely careful drafting around sensitive data, under-extraction of sensitive-data claims, or a category definition that routes such statements into FIRST\_PARTY), and a targeted audit of sensitive-data language would be needed to resolve this. The same applies to AUTOMATED\_DECISIONS (40 OPPT statements, 1 judged pair, 0 confirmed), a category of growing regulatory interest given Article 22 of the GDPR~\citep{gdpr2016}.

We detect textual contradictions but do not test whether users are actually affected. Behavioral studies measuring whether reading a commitment leads to expectations that documented practices contradict are needed to establish real-world impact.

We do not distinguish between intentional and structural contradictions, nor do we filter based on legal defensibility. Contradictions involving regulatory exceptions (M\&A clauses, HIPAA permissions, law enforcement requirements) remain valid findings: a policy stating ``we obtain authorization'' followed by ``we do not need authorization for [exception]'' contains a textual contradiction regardless of legal permissibility.

The analysis is limited to English-language policies, and policies in other languages may exhibit different contradiction patterns. Coverage is also limited jurisdictionally: although some analyzed policies are partly governed by the GDPR, we do not analyze policies drafted primarily for non-U.S.\ regulatory frameworks such as the GDPR or Brazil's Lei Geral de Prote\c{c}\~{a}o de Dados (LGPD)~\citep{lgpd2018}. Policies were collected at specific dates and may have been updated since collection.

\section{Future Work}
\label{sec:future}

Several directions emerge from these findings; we order them by priority.

The stability experiment (Section~\ref{sec:stability}) performed three of the inexpensive methodological experiments identified in earlier sections: judging the NLI-flagged pairs below the similarity threshold, separating the judge panel from the extraction panel, and placing the cross-corpus comparison on a matched filter configuration. Two remain. Re-running the judge stage several times on identical pairs would measure verdict stability under nominally deterministic decoding, isolating the judge-stage contribution that the stability run (which changed extraction and judging together) cannot. And disabling the qualifier-coverage filter branch implicated by the legacy-pair audit (Section~\ref{sec:metadata}), then judging the pairs it discards, remains the highest-value single ablation, since it targets the one filter that within-corpus evidence indicts; the stability run retained that filter.

Human expert validation is the priority for peer-reviewed publication. A rigorous validation study would involve sampling statement pairs stratified by verdict type, independent labeling by privacy law experts blind to LLM verdicts, and computing precision and recall against this human gold standard. Short of full expert panels, a minimum viable design, two trained annotators labeling a stratified sample of roughly 100 pairs against a written codebook, would yield a usable precision estimate with binomial uncertainty and is feasible for a single research group. Behavioral studies testing whether users who read a commitment form expectations that documented practices contradict would quantify the real-world impact of privacy washing.

Temporal analysis tracking how contradictions evolve as companies update policies would reveal whether contradictions are resolved, persist, or accumulate over time. Applying the pipeline to current (2025--2026) Fortune 500 policies would provide a large-scale contemporary measurement, and existing longitudinal and large-scale policy corpora, such as the million-document historical dataset of \citet{amos2021privacy} and the PrivaSeer corpus~\citep{srinath2021corpus}, offer natural targets for scaling the analysis. Integration with code-versus-policy analysis~\citep{andow2020policheck} would enable end-to-end auditing from policy text to app behavior.

Remediation tools that suggest specific edits to resolve detected contradictions could transform the pipeline from a diagnostic to a prescriptive instrument. The healthcare-specific patterns observed in Kaleida Health (HIPAA authorization requirements contradicted by legally-permitted exceptions) suggest that sector-specific analysis may reveal distinct contradiction structures. OPP-115 contains only two dedicated healthcare facilities (Kaleida Health and GW Medical Faculty Associates), insufficient for robust sector-wide analysis. Future work with dedicated healthcare corpus sampling (20--50 hospitals, clinics, and insurers) would be necessary to characterize whether healthcare privacy washing follows generalizable patterns distinct from technology-sector findings. Extension to non-English policies would test whether the patterns we observe are specific to U.S.\ regulatory contexts or reflect universal structural tendencies in privacy policy composition.

\section{Responsible Disclosure and Ethical Considerations}
\label{sec:ethics}

This paper names specific companies as containing panel-confirmed contradictions. We report company names because the findings are only verifiable against the public source policies if the companies are identified, and because anonymization would prevent the community validation we invite. Three considerations accompany this choice.

First, the verdicts are LLM-generated and have not been validated by human experts; precision is unknown (Section~\ref{sec:limitations}). Any individual finding may be a false positive. The findings are not allegations of legal violation or of intentional deception: they identify textual conditions (a commitment and a practice in pragmatic tension within one public document) that warrant review, and both statements in every pair are quoted from or traceable to the company's own published policy. Companies named in this paper were not notified before its publication; the analyzed texts are public documents, and the findings assert nothing beyond what those documents state. We judged pre-publication notification unnecessary for this preprint because the findings are unvalidated candidates drawn entirely from public documents and framed as such; if precision is established through expert validation and findings are reframed as compliance signals, advance notification with a response window would be appropriate before venue submission. We welcome corrections from named companies and will note any in revised versions.

Second, the three judge models are produced by Anthropic, OpenAI, and Google, and all three companies appear in the OPPT corpus. Judge models therefore evaluated statement pairs drawn from their own providers' privacy policies. This is a conflict of interest inherent to using frontier LLMs to analyze the policies of frontier LLM providers. We examined the per-judge verdicts on these pairs (all included in the released artifacts). Google had 4 judged pairs, all confirmed (three are legacy pairs; one survives in the reproducible subset); the Google judge dissented on 2 of the 4, but its dissent rate on confirmed pairs overall is 48\% (16 of 33), so this is indistinguishable from its general conservatism (Table~\ref{tab:agreement}). Anthropic and OpenAI each had one judged pair, unanimously rejected by all three judges, including the non-self judges; these samples are too small to be informative. We find no evidence of provider self-preference, while noting the analysis has essentially no statistical power at these sample sizes. Findings involving these three companies should be read with this conflict in mind. The stability run (Section~\ref{sec:stability}) removes the conflict entirely by judging with models from providers absent from both corpora; under that panel, none of the Google confirmations recur, consistent with the legacy-pair provenance of three of the four (the stability run retains the filter that excludes that pair class, so this non-recurrence is partly by construction).

Third, the term ``privacy washing'' carries connotations of deliberate misconduct that our structural findings do not establish. We have used the term for its continuity with the greenwashing literature's structural tradition (Section~\ref{sec:definition}) and repeat here that no finding in this paper attributes motive to any policy author.

The analysis uses only publicly available policy documents; no personal data was collected or processed.

\section{Conclusion}

We introduced \emph{privacy washing}, the presence of internal contradictions within privacy policies where commitments are undermined by documented practices, and measured it across two corpora collected 11 years apart. Our four-stage pipeline identified panel-confirmed contradictions in 12.2\% of OPPT companies as run (9.8\% in the reproducible subset that excludes pairs carried over from an earlier pair-generation pass) and 36.5\% of OPP-115 companies; the two runs used different filter configurations, so the cross-corpus difference is not interpretable as an era effect. Contradictions involving third-party sharing account for the majority of confirmed cases in both primary runs, and the recurrence of the same category patterns across regulatory eras is consistent with structural factors in policy composition. A stability re-run with separated extraction and judge panels and matched filter configurations (Section~\ref{sec:stability}) reproduced the OPPT prevalence under the original protocol, did not reproduce the legacy-pair findings (partly by its own design), measured substantial confirmable recall below the similarity threshold, and showed that the third-party majority, though not the recurrence of the underlying category pairs, is panel-sensitive.

The single measurement that would most change this picture is expert validation: no reported quantity currently has a known relationship to human judgment, and blind annotation by privacy law experts would convert panel-confirmed counts into precision and recall estimates. Until then, the pipeline is best understood as a candidate-generation tool whose practical utility depends on precision that remains to be established; it is not a measurement instrument or an enforcement tool. For regulators, it can prioritize policies for expert scrutiny under the FTC's net impression doctrine; for companies, it can surface candidate inconsistencies before publication; for researchers, the concept and methodology open policy quality assessment beyond readability. We release the pipeline, intermediate outputs, and judge results to enable community validation and extension.

\section*{Competing Interests}

The author is the founder of Varitas, a company developing privacy policy analysis technology related to the pipeline described in this paper. No external funding supported this work.

\section*{Data and Code Availability}

The detection pipeline, all intermediate outputs (extracted statements, NLI scores, judge verdicts), and analysis scripts are available at \url{https://github.com/Varitas-Foundation/privacy-washing}. The three-LLM consensus panel used for both extraction and judging in the primary runs comprises \texttt{anthropic/claude-haiku-4.5}, \texttt{openai/gpt-5-mini}, and \texttt{google/gemini-3-flash-preview} via OpenRouter; the primary experiment runs were executed on January 31, 2026 (OPPT) and February 3, 2026 (OPP-115). The stability run (Section~\ref{sec:stability}), executed August 30--31, 2026, extracts with \texttt{anthropic/claude-haiku-4.5}, \texttt{openai/gpt-5.6-luna}, and \texttt{google/gemini-3.7-flash} and judges with \texttt{deepseek/deepseek-v4-flash-0731}, \texttt{z-ai/glm-5.3-flash}, and \texttt{moonshotai/kimi-k3}; its artifacts, including a discarded first judge pass in which one judge model was unreachable under the account's data-routing policy, are released in the same repository. The judge prompt appears in Appendix~\ref{appendix:judge}; complete response logs, including per-judge verdicts on pairs from the judge providers' own companies (see Section~\ref{sec:ethics}), are included in the repository. The OPPT corpus annotations are released under CC-BY-4.0. OPP-115 data is available from its original source~\citep{wilson2016creation}. A versioned archival snapshot of the repository will accompany the final version.

\appendix
\section{LLM Judge Prompt}
\label{appendix:judge}

The following prompt was used for all three LLM judges (Claude Haiku 4.5, GPT-5 mini, Gemini 3 Flash Preview) to evaluate statement pairs. It is reproduced in full, reformatted for LaTeX, from \texttt{scripts/judge\_statement\_prompt.md} in the released repository; the prompt is unversioned in the repository and was not modified between runs. Each judge received the same prompt with company name and statement text substituted into the template variables.

\begin{quote}
\small
You are a privacy policy analyst assessing whether two atomic statements from the same company's privacy policy contradict each other.

\medskip
\textsc{Statement A (Commitment):} A specific promise, reassurance, or limitation the company places on itself regarding data handling.

\medskip
\textsc{Statement B (Practice):} A description of what the company actually does with data.

\medskip
\textsc{Your Task:} Determine whether Statement B contradicts the commitment made in Statement A.

\medskip
\textsc{What Is a Contradiction:}
\begin{itemize}
\item Statement A makes a specific promise that Statement B violates or undermines
\item Example: A: ``We do not sell your personal information.'' / B: ``We share data with advertising partners for targeted marketing.'' $\to$ Contradiction
\item Example: A: ``We limit data sharing to essential service providers.'' / B: ``We share user data with affiliates, partners, and advertisers.'' $\to$ Contradiction
\end{itemize}

\medskip
\textsc{What Is Not a Contradiction:}
\begin{itemize}
\item B implements or supports what A promises (e.g., A: ``We protect your data'' / B: ``We use industry-standard encryption'' $\to$ not a contradiction)
\item B restates the same commitment in different words
\item B describes an unrelated practice with no bearing on A's commitment
\item B is about a different data type, user group, or context than A
\item A uses hedging language (``may'', ``might'') and B describes that possibility occurring
\item A and B are about different products or services within the same company
\end{itemize}

\medskip
\textsc{Examples:}

\medskip
Example 1 (CONTRADICTION). Commitment: ``We do not share personal information with third parties for marketing.'' Practice: ``We provide user data to advertising partners to deliver personalized ads.'' Verdict: CONTRADICTION; the commitment prohibits marketing sharing, but the practice describes sharing with advertising partners.

\medskip
Example 2 (NOT\_CONTRADICTION). Commitment: ``We implement security measures to protect your personal information.'' Practice: ``We collect device identifiers and IP addresses for service improvement.'' Verdict: NOT\_CONTRADICTION; the practice describes data collection, which is unrelated to the security commitment. Different topics, no conflict.

\medskip
Example 3 (CONTRADICTION). Commitment: ``We do not sell personal information of consumers under 16.'' Practice: ``Service providers acting on our behalf may process and sell user data.'' Verdict: CONTRADICTION; the commitment claims no selling, but the practice describes service providers who may sell data.

\medskip
Example 4 (NOT\_CONTRADICTION). Commitment: ``We do not sell your personal data.'' Practice: ``We collect browsing history and usage data to improve our services.'' Verdict: NOT\_CONTRADICTION; collecting data for internal improvement is distinct from selling it. No conflict between these specific claims.

\medskip
Respond with ONLY this JSON (no other text): \texttt{\{``verdict'': ``CONTRADICTION'' or ``NOT\_CONTRADICTION'', ``reasoning'': ``1-2 sentence explanation''\}}
\end{quote}

\section{Statement Extraction Prompt}
\label{appendix:extraction}

The following prompt (version~2, from \texttt{scripts/statement\_extraction\_prompt\_v2.md} in the released repository) was used by all three extraction models. Template variables supply the segment text, company name, category, and, where available, the guiding annotation. Statements are retained when extracted by at least two of the three models, matched by cosine similarity of sentence embeddings at threshold 0.7.

\begin{quote}
\small
You are extracting atomic statements from a privacy policy segment.

\medskip
\textsc{Segment Text:} \{segment\_text\} \quad \textsc{Company:} \{company\} \quad \textsc{Category:} \{category\} \quad \{annotation\_block\}

\medskip
\textsc{Task:} Extract each distinct claim or practice as a separate atomic statement with enhanced metadata. For each statement, provide:

\begin{enumerate}
\item \emph{text}: A self-contained sentence (10--30 words) that captures one specific claim or practice. Include enough context to be understood without the original segment.
\item \emph{type}: COMMITMENT (a promise, reassurance, or limitation the company places on itself: ``we do not\ldots'', ``we limit\ldots'', ``we protect\ldots'') or PRACTICE (a description of what the company actually does with data: ``we collect\ldots'', ``we share\ldots'', ``we use\ldots'').
\item \emph{subject}: WHO performs the action: COMPANY (``we collect'', ``we share''), SERVICE\_PROVIDER (``our processors''), THIRD\_PARTY (``advertisers may''), AFFILIATES (``our subsidiaries''), or USER (``you can opt out'').
\item \emph{aspect}: WHAT data lifecycle stage: COLLECTION, USE, SHARING, SALE, RETENTION, DELETION, ACCESS\_CONTROL, or SECURITY.
\item \emph{scope}: Under WHAT conditions: UNIVERSAL, CONDITIONAL, CONSENT\_BASED, LEGAL\_REQUIREMENT (M\&A, subpoenas, law enforcement), or GEOGRAPHIC\_LIMITED (California, EU residents, etc.).
\item \emph{qualifiers}: Array of key limiting phrases verbatim from source text (e.g., ``except as required by law'', ``only for service delivery''); empty array if none.
\item \emph{category}: The OPPT category this statement relates to (e.g., FIRST\_PARTY, THIRD\_PARTY, TRACKING, SALE\_SHARING, SECURITY).
\end{enumerate}

\medskip
\textsc{Rules:} Each statement must be self-contained (understandable without the original paragraph). Preserve the original meaning; do not soften or strengthen language. If a sentence contains both a commitment and a practice, split into two statements. Preserve qualifiers exactly as written in the source text. Ignore procedural/navigational text (``click here'', ``see our policy''). If the segment is purely informational with no claims or practices, return an empty list.

\medskip
\textsc{Examples} (input segment $\to$ JSON output, reproduced verbatim):

\medskip
Input: ``We do not sell your personal information, except as required by law or in connection with a merger or acquisition.'' $\to$ \texttt{\{"text": "The company does not sell personal information.", "type": "COMMITMENT", "subject": "COMPANY", "aspect": "SALE", "scope": "UNIVERSAL", "qualifiers": ["except as required by law", "in connection with a merger or acquisition"], "category": "SALE\_SHARING"\}}

\medskip
Input: ``Third-party advertisers may collect information about your browsing activity across websites.'' $\to$ \texttt{\{"text": "Third-party advertisers collect information about user browsing activity across websites.", "type": "PRACTICE", "subject": "THIRD\_PARTY", "aspect": "COLLECTION", "scope": "UNIVERSAL", "qualifiers": [], "category": "TRACKING"\}}

\medskip
Input: ``With your consent, we share your location data with our advertising partners.'' $\to$ \texttt{\{"text": "The company shares user location data with advertising partners when the user provides consent.", "type": "PRACTICE", "subject": "COMPANY", "aspect": "SHARING", "scope": "CONSENT\_BASED", "qualifiers": ["with your consent"], "category": "THIRD\_PARTY"\}}

\medskip
The prompt closes with a strict JSON-only output template.
\end{quote}

\bibliography{references}

\end{document}